\documentclass{jfm}
\usepackage{amsmath}  
\usepackage{amssymb}  
\usepackage{mathtools}
\usepackage{placeins}
\usepackage{tikz}
\newcommand*\circled[1]{\tikz[baseline=(char.base)]{
    \node[shape=circle,draw,inner sep=1.3pt] (char) {#1};}}
\usepackage{graphicx}
\usepackage{newtxtext}
\usepackage{newtxmath}
\usepackage{natbib}
\usepackage{hyperref}
\usepackage{bm}

\DeclareMathOperator{\sech}{sech}
\hypersetup{
    colorlinks = true,
    urlcolor   = blue,
    citecolor  = black,
}

\newcommand{\RomanNumeralCaps}[1]
\linenumbers

\title{Bidirectional shallow-water wave turbulence}

\author{Ashleigh Simonis\aff{1}
  \corresp{\email{asimonis@umich.edu}},
  Sergey Nazarenko\aff{2},
  Jalal Shatah\aff{3}
 \and Yulin Pan\aff{1}\corresp{\email{yulinpan@umich.edu}}}

\affiliation{\aff{1}Department of Naval Architecture and Marine Engineering, University of Michigan, 2600 Draper Drive, Ann Arbor, Michigan 48109, USA
\aff{2}Institut de Physique de Nice, Université Côte d'Azur et CNRS, 17 rue Julien Lauprêtre, 06200 Nice, France
\aff{3}Courant Institute of Mathematical Sciences, New York University, 251 Mercer Street, New York, New York 10012, USA}

\begin{document}
\maketitle

\begin{abstract}
We study bidirectional one-dimensional (1-D) shallow-water waves within a class of Boussinesq equations, including the integrable Kaup-Boussinesq (KB) equation and a truncated-dispersion variant, which serves as a representative non-integrable model. For these two systems, the normal-form transformation yields an interaction coefficient of the same general structure, differing only through the dispersion relation. We derive this coefficient and numerically confirm that it vanishes on the resonant manifold for the KB equation, as expected in the literature. In contrast, the non-integrable model admits a non-vanishing interaction coefficient, producing a non-trivial wave kinetic equation (WKE), which is the first known in a 1-D shallow-water setting. The resulting WKE is non-homogeneous in nature due to the non-homogeneity of the corresponding dispersion relation; however, approximate Kolomogrov-Zakharov (KZ) solutions can be derived in a novel way under certain approximations. Numerical experiments in two settings validate the kinetic predictions and elucidate the underlying dynamics: (i) in free-evolution cases of the KB equation, despite complete integrability and the invariance of the discrete nonlinear spectrum guaranteed by isospectrality, an initial arbitrary wavenumber spectrum undergoes substantial evolution driven by quasi-resonant triad interactions; (ii) in forced-dissipated cases of the non-integrable equation, we find stationary power-law spectra that agree with the theoretical predictions.
\end{abstract}


\section{\label{sec:introduction}Introduction}

Wave kinetic theory describes the long-time statistical evolution of weakly nonlinear dispersive waves, predicting irreversible spectral transfer toward either a thermal equilibrium state or a Kolmogorov-Zakharov cascade \citep[e.g.,][]{Zakharov2012,Nazarenko2011}. While this theory is well developed for deep-water waves, beginning with Hasselmann’s derivation of the wave kinetic equation \citep{Hasselmann1962}, its shallow-water counterpart is less understood in the literature. For 2-D shallow-water surface waves, the theory has been studied starting from either the Euler equations \citep{Zakharov1999} or the Boussinesq equations \citep{Onorato2009}, with their predictions of the KZ spectrum observed in wave tank experiments \citep{Kaihatu2007}. However, in the 1-D setting, there is currently no established theory for wave turbulence, as many contend the dynamics can be described by integrable models \citep{Zakharov2012}, such as the Korteweg-de Vries (KdV), Kadomtsev-Petviashvili-2 (KP-2), Camassa-Holm (CH), and Kaup-Boussinesq (KB) equations. This line of reasoning is due in part to the widely held empirical belief that integrable systems do not admit non-trivial kinetic descriptions \citep[e.g.,][]{Zakharov1980,Zakharov1988,Zakharov1991}.


The purpose of this paper is to revisit the largely underexplored problem of wave turbulence in bidirectional 1-D shallow-water waves, starting from the $abcd$ class of Boussinesq equations, which provide descriptions of the dynamics in the long-wave regime. This class includes the completely integrable KB equation \citep[e.g.,][]{Kaup1975,Kupershmidt1985,El2001,Ivanov2009,Bhrawy2013,Gong2022,Klein2025}, as well as several well-known non-integrable models such as the classical Boussinesq, Bona-Smith, and Benjamin-Bona-Mahony (BBM) systems \citep[e.g.,][]{Bona2002,Bona2004}. From an experimental perspective, studies of bidirectional shallow-water waves \citep[e.g.,][]{Redor2019,Redor2021,Leduque2025} have demonstrated a range of behaviours, including spectral evolution towards a flat energy spectrum at large scales (interpreted as thermalisation in \citet{Pelinovsky2006}) and/or the emergence of a soliton gas, depending on the system nonlinearity and forcing condition. These observations suggest that it is problematic to view either the completely integrable KB equation or standard non-integrable Boussinesq equations as a standalone, canonical model for 1-D bidirectional shallow-water waves, as each only accounts for a subset of the results, i.e., integrable features in the former and more traditional wave kinetic behaviour in the latter.

To develop a comprehensive understanding of both integrable and non-integrable dynamics, we consider the completely integrable KB equation and a truncated-dispersion variant representing the non-integrable setting. Although exact dispersion relations differ across the $abcd$ class, this truncation retains the common leading-order long-wave dispersive behaviour. A normal-form transformation can be applied to the models considered to remove the quadratic terms, for which we carefully characterise its validity condition depending on the parameters $\alpha$ and $\beta$ measuring the nonlinearity and dispersion. This yields an interaction coefficient of the same general form for both cases, differing only in the dispersion relation. For the KB equation, we numerically confirm that the interaction coefficient is zero exactly on the four-wave resonant manifold, consistent with the prevailing view in the literature \citep[e.g.,][]{Zakharov1980,Zakharov1988,Zakharov1991}. In contrast, when considering the truncated-dispersion model, our analysis yields a non-zero interaction coefficient that can be reduced to a compact form, clearly demonstrating non-trivial wave kinetics. The resulting WKE is inherently non-homogeneous due to the inhomogeneity of the dispersion relation. While the thermal equilibrium spectra remain as exact stationary solutions, the KZ solutions must be derived under approximations in a novel way. We derive two distinct direct cascade solutions corresponding to the limits of small and large modal amplitudes at low wavenumbers, given respectively by $n_k\sim |k|^{-3}$ and $n_k\sim |k|^{-3/2}$, with each corresponding to a local solution of the WKE.

We proceed to perform direct numerical simulations, both to test the validity of our theoretical predictions and to investigate the underlying mechanism of energy transfer. We start by demonstrating a free-evolution case of the KB equation from random-wave initial conditions, where the discrete eigenvalues obtained from the direct scattering transform (representing solitons present in the initial field) remain constant in time, while the wavenumber spectrum quickly evolves towards a thermalised state. This case demonstrates that, despite the KB equation being completely integrable and maintaining isospectrality of the eigenvalues, it can still exhibit significant spectral evolution due to quasi-resonant triad interactions. We then turn to simulations of the forced-dissipated non-integrable Boussinesq equation, where the low-wavenumber modal amplitude is controlled through the presence of the low-wavenumber dissipation. For small and large modal amplitudes at low wavenumbers, respectively, we reproduce the two theoretical KZ solutions. We also discuss how wave kinetic behaviour may become possible even in the full KB model as a result of four-wave quasi-resonant interactions. This is due to the previously unidentified near-singular behaviour of the interaction coefficient near the resonant manifold. A more detailed study is left for future work.

We finally remark that the wave kinetic behaviour we discover in the present paper is fundamentally different from Janssen's WKE for the 1-D Nonlinear Schr{\"o}dinger (NLS) equation \citep{Janssen2003}, as well as the notion of ``integrable turbulence" \citep[e.g.,][]{Agafontsev2015,Agafontsev2016,Randoux2016,Roberti2019,El2021,Bonnemain2022,Suret2024}. The former describes a time-dependent collision integral that vanishes as $t\rightarrow\infty$ (thus saturating to a steady state spectrum), whereas the latter may refer to a soliton-gas state in an integrable model, described statistically by a kinetic theory for the ensemble of solitons. It is often assumed that the soliton-gas state is the only physically relevant phenomenon for 1-D shallow water waves. Our results demonstrate that this is not the case, as free-wave, non-resonant interactions and spectral transfer may occur in a more traditional wave-turbulence sense, providing insight into integrable systems in a broader context. Taken together with the non-integrable Boussinesq case, which supports true resonant interactions and a non-trivial WKE, these results establish a robust theoretical framework for 1-D shallow-wave wave turbulence.

\section{\label{sec:theoretical-deriv}Theoretical derivation}
\subsection{Derivation of the wave kinetic equation}\label{wke-nf}
In this section, we use the Kaup-Boussinesq (KB) equation as a starting point to develop a generalised framework that we apply to two representative cases within the $abcd$ class of Boussinesq equations. The resulting normal-form interaction coefficient has the same general structure in these cases, with the dispersion relation being the only model-dependent component. We will examine the interaction coefficient for both the full KB variant and its truncated-dispersion version.

The dimensional (KB) system is given by
\label{derivation}
\begin{equation}
    \eta_t + [(h+\eta)u]_x=-\frac{h^3}{3}u_{xxx},
\label{eq:kb1}
\end{equation}
\begin{equation}
    u_t+uu_x+g\eta_x=0,
\label{eq:kb2}
\end{equation}
where $\eta(x,t)$ denotes the surface elevation, $u(x,t)$ is the fluid velocity, $g$ is the acceleration of gravity, and $h$ is the water depth. The system of \eqref{eq:kb1} and \eqref{eq:kb2} is known to be completely integrable in the inverse scattering method (ISM) sense \citep{Kaup1975}. We introduce the following non-dimensional variables:
\begin{equation}
\begin{gathered}
    \tilde{\eta} = \eta / a, \quad \tilde{x} = x / L_p,\quad \tilde{t} = t / (L_p / \sqrt{gh}), \quad \tilde{u} = u /(\sqrt{gh} \frac{a}{h}).
    \label{eq:nond}
\end{gathered}
\end{equation}
where $a$ denotes the characteristic wave amplitude (e.g., half the significant wave height $H_s$) and $L_p$ is the peak wavelength of the wave field.  Substituting \eqref{eq:nond} into \eqref{eq:kb1} and \eqref{eq:kb2} and omitting the tilde notation, we obtain
\begin{equation}
    \eta_t + u_x + \alpha(\eta u)_x = -\frac{1}{3}\beta u_{xxx},
\label{eq:kb_n1}
\end{equation}
\begin{equation}
    u_t+\eta_x+\alpha uu_x=0,
\label{eq:kb_n2}
\end{equation}
where $\alpha = a/h$ and $\beta = (h/L_p)^2$ are dimensionless measures of nonlinearity and dispersion, respectively.

We consider \eqref{eq:kb_n1} and \eqref{eq:kb_n2} on a periodic domain $\mathbb{T}_L$. Defining Fourier-domain variables such as $\hat{\eta}_k(t)=\int_0^L \eta(x,t) e^{-ikx} dx/L$ (note that $k$ can be positive or negative), \eqref{eq:kb_n1} and \eqref{eq:kb_n2} can be transformed into:
\begin{equation}
    \frac{\partial \hat{\eta}_k}{\partial t} + i(k-\frac{1}{3}\beta k^3) \hat{u}_k + i\alpha \sum_{k_1+k_2=k} k \hat{\eta}_{k_1} \hat{u}_{k_2} = 0,
\label{eq:kb_F1}
\end{equation}
\begin{equation}
    \frac{\partial \hat{u}_k}{\partial t} + ik \hat{\eta}_k + \frac{i}{2} \alpha \sum_{k_1+k_2=k} k  \hat{u}_{k_1} \hat{u}_{k_2} =0.
\label{eq:kb_F2}
\end{equation}
The linear part of \eqref{eq:kb_F1} and \eqref{eq:kb_F2} yields the dispersion relation
\begin{equation}
   \omega_k = \kappa \sqrt{1-\frac{1}{3}\beta \kappa^2}.
\label{eq:disp}
\end{equation}
where $\kappa = |k|$. Assuming $\beta \kappa^2 \ll 1$, we expand in the small parameter $\beta\kappa^2$ and retain terms through first order
\begin{equation}
    \omega_k=\kappa\left(1-\frac{\beta}{6}\kappa^2\right),
    \label{eq:disp-trunc}
\end{equation}
which we use to define the truncated-dispersion model below. This truncated form captures the leading-order dispersive behaviour of non-integrable models in the $abcd$ class \citep[e.g.,][]{Bona2002,Bona2004}. Both \eqref{eq:disp} and \eqref{eq:disp-trunc} agree with the shallow-water dispersion relation $\omega_k=\kappa$ when $\beta\kappa^2\ll1$, but depart significantly from it on the descending branch,
\begin{equation}
    \kappa > \kappa_*,\qquad 
    \frac{d\omega}{d\kappa}\Big|_{\kappa=\kappa_*}=0,
    \label{eq:desc_branch}
\end{equation}
where $\kappa_*=\sqrt{3/(2\beta)}$ for the full KB equation and $\kappa_*=\sqrt{2/\beta}$ for the truncated variant. In practice, we limit to a finite wavenumber range $\kappa_{\mathrm{max}}<\kappa_*$, omitting the descending branch to ensure we remain in the near-shallow water regime for the subsequent theoretical and numerical analyses.
 
By defining the canonical variables,
\begin{equation}
    \hat{\eta}_k = \frac{\sqrt{2}}{2}\omega_k^{1/2} (a_k + a_{-k}^*)
\label{eq:can_eta}
\end{equation}
and
\begin{equation}
    \hat{u}_k = \frac{\sqrt{2}}{2}\frac{k}{\omega_k^{1/2}} (a_k - a_{-k}^*),
\label{eq:can_u}
\end{equation}
leading to
\begin{equation}
    a_k = \frac{\sqrt{2}}{2}\omega_k^{-1/2}\hat{\eta}_k + \frac{\sqrt{2}}{2} \frac{\omega_k^{1/2}}{k}\hat{u}_k,
\label{eq:can_a}
\end{equation}
we can diagonalise \eqref{eq:kb_F1} and \eqref{eq:kb_F2} in the variable $a_k$
\begin{equation}
\begin{split}
        \frac{\partial a_k}{\partial t} + i\omega_k a_k+ \frac{i\alpha}{4\sqrt{2}} &\sum_{k_1+k_2=k}\{ U_{k12}a_{k_1}a_{k_2} - U_{k12}a_{k_1}a_{-k_2}^*\\
        &+S_{k12}a_{-k_1}^*a_{k_2} - S_{k12} a_{-k_1}^*a_{-k_2}^*\} = 0,
\end{split}
\label{eq:aequation}
\end{equation}
where
\begin{equation}
    U_{k12}= [2W(k,2)+W(1,2)] (\omega_{k}\omega_{k_1}\omega_{k_2})^{1/2},
\end{equation}
\begin{equation}
    S_{k12}= [2W(k,2)-W(1,2)] (\omega_{k}\omega_{k_1}\omega_{k_2})^{1/2},
\end{equation}
with 
\begin{equation}
\begin{aligned}
W(i,j)=\frac{k_ik_j}{\omega_i\omega_j}.
\end{aligned}
\end{equation}

Manipulating \eqref{eq:aequation} using symmetry conditions, we get the final canonical equation
\begin{equation}
\begin{split}
\frac{\partial a_k}{\partial t} + i\omega_k a_k + &i \{\sum_{k_1+k_2=k} V_{k12} a_{k_1}a_{k_2} + \sum_{k_1-k_2=k} V_{k12} a_{k_1}a_{k_2}^*\\
&+ \sum_{k_2-k_1=k} V_{k12} a_{k_1}^*a_{k_2} + \sum_{k_2+k_1+k=0} V_{k12} a_{k_1}^*a_{k_2}^*
    \} = 0,
\label{eq:a_sym}
\end{split}
\end{equation}
where
\begin{equation}
\begin{split}
    V_{k12}=&\frac{\alpha(\omega_{k}\omega_{k_1}\omega_{k_2})^{1/2}}{4\sqrt{2}}[W(k,1)+W(k,2)+W(1,2)].
    \label{Vkernel}
\end{split}
\end{equation}
We note that all interaction kernels in \eqref{eq:a_sym} are the same, which is consistent with the case in the 2-D Boussinesq equation derived in \citet{Onorato2009}.

Our next goal is to apply a normal-form transformation (or multiple time scale method) to \eqref{eq:a_sym} with the aim of eliminating the quadratic nonlinear terms, since the dispersion relation \eqref{eq:disp} does not support triad resonant interactions. This approach has been previously applied to the 2-D Euler equations for shallow water \citep{Zakharov1999} and to the 2-D Boussinesq equations \citep{Onorato2009}. The procedure is standard \citep[e.g.,][]{Krasitskii1994,Zakharov2012} except for the validity condition of the transformation, which will be discussed later.

We proceed by applying the normal-form transformation method to introduce a new variable $b_k$ defined as
\begin{equation}
\begin{split}
    a_k=\ b_k&+ \sum_{k_1+k_2=k} \Gamma^{[1]}_{k12}b_{k_1}b_{k_2}- 2 \sum_{k_2-k_1=k} \Gamma^{[1]}_{2k1}b_{k_1}^*b_{k_2}+ \sum_{k_1+k_2+k=0} \Gamma^{[2]}_{k12}b_{k_1}^*b_{k_2}^*\\
    &+\sum_{k+k_1=k_2+k_3} B_{0123} b_{k_1}^*b_{k_2}b_{k_3}+ \text{other cubic terms},
\end{split}
    \label{akbk}
\end{equation}
such that rewriting \eqref{eq:a_sym} in terms of $b_k$ yields
\begin{equation}
   \frac{\partial b_k}{\partial t} + i\omega_k b_k + i\sum_{k+k_1=k_2+k_3} T_{k123} b_{k_1}^*b_{k_2}b_{k_3} = 0.
    \label{eq:bk}
\end{equation}
This transformation is achieved by appropriately choosing the coefficients in \eqref{akbk}, for example,
\[
   \Gamma^{[1]}_{k12} = \frac{-V_{k12}}{\omega_k-\omega_1-\omega_2},\quad\Gamma^{[2]}_{k12} = \frac{-V_{k12}}{\omega_k+\omega_1+\omega_2}.
\]
\\
The interaction kernel in \eqref{eq:bk} is given by

\begin{equation}
\begin{split}
    T_{k123} =&-V_{k,2,k-2}V_{3,1,3-1}\Big[\frac{1}{\omega_2+\omega_{k-2}-\omega_k} + \frac{1}{\omega_1+\omega_{3-1}-\omega_3} \Big] \\
               & -V_{1,2,1-2}V_{3,k,3-k}\Big[\frac{1}{\omega_2+\omega_{1-2}-\omega_1} + \frac{1}{\omega_k+\omega_{3-k}-\omega_3} \Big] \\
               & -V_{k,3,k-3}V_{2,1,2-1}\Big[\frac{1}{\omega_3+\omega_{k-3}-\omega_k} + \frac{1}{\omega_1+\omega_{2-1}-\omega_2} \Big] \\
               & -V_{1,3,1-3}V_{2,k,2-k}\Big[\frac{1}{\omega_3+\omega_{1-3}-\omega_1} + \frac{1}{\omega_k+\omega_{2-k}-\omega_2} \Big] \\
               & -V_{k+1,k,1}V_{2+3,2,3}\Big[\frac{1}{\omega_{k+1}-\omega_k-\omega_1} + \frac{1}{\omega_{2+3}-\omega_2-\omega_3} \Big] \\
               & -V_{-k-1,k,1}V_{-2-3,2,3}\Big[\frac{1}{\omega_{k+1}+\omega_k+\omega_1} + \frac{1}{\omega_{2+3}+\omega_2+\omega_3} \Big].
\end{split}
    \label{Tkernek}
\end{equation}

We observe that $T_{k123}$ in \eqref{Tkernek} possesses the desired symmetry properties, namely, $T_{k123}=T_{1k23}=T_{k132}=T_{23k1}$.

To obtain the above results, the transformation $a_k \rightarrow b_k$ defined in \eqref{akbk} must be near-identity. As discussed in \citet{Zakharov1999}, this requires
\begin{equation}
\begin{split}
    \Big|\sum_{k_1+k_2=k} \Gamma^{[1]}_{k12}b_{k_1}b_{k_2} - 2 \sum_{k_2-k_1=k} \Gamma^{[1]}_{2k1}b_{k_1}^*b_{k_2}+ \sum_{k_1+k_2+k=0} \Gamma^{[2]}_{k12}b_{k_1}^*b_{k_2}^* \Big| \ll |b_k|
    \label{nearI}
\end{split}
\end{equation}
In formulating \eqref{nearI}, inconsistencies arise in \citet{Zakharov1999}. In particular, the analysis in \citet{Zakharov1999} does not account for the random phases of $b_k$ when estimating the summations on the left-hand side, resulting in an overestimation. In Appendix \ref{appA}, we provide a new estimation that properly incorporates phase randomness to derive the validity condition for the normal-form transformation (see also \citep{Wu2025}). The result is straightforward and can be summarised as follows: 

\emph{We consider a spectrum with finite bandwidth $\Delta$ in $k$, assuming that it decays sufficiently fast outside $\Delta$. Then \eqref{nearI} corresponds to the condition of $\alpha \ll \beta$.} While the detailed derivation is provided in Appendix \ref{appA}, the underlying logic is to avoid small denominators arising on the left-hand side of \eqref{nearI}. The most singular term in the denominator is of $O(\beta)$, which must be significantly larger than the numerator of $O(\alpha)$.   

Starting from \eqref{eq:bk}, we formally derive the corresponding wave kinetic equation (WKE) following standard procedures \citep[e.g.,][]{Nazarenko2011}. The WKE is given by
\begin{equation}
\begin{split}
    \frac{\partial n_k}{\partial t} =\ &4\pi \int |T_{k123}|^2 n_kn_1n_2n_3 \left[ \frac{1}{n_k}+\frac{1}{n_1}-\frac{1}{n_2}-\frac{1}{n_3} \right]\\
    &\times\delta(k+k_1-k_2-k_3)\delta(\omega_k+\omega_1-\omega_2-\omega_3) dk_1dk_2dk_3,
    \label{WKE}
\end{split}    
\end{equation}
where $n_k= \langle b_k b_k^* \rangle \approx \langle a_k a_k^* \rangle$ denotes the wave action. 

To determine whether \eqref{WKE} is trivial or not, it is essential to analyse the interaction coefficient $T_{k123}$ on the resonant manifold defined by the resonance condition in \eqref{WKE}. We begin by emphasising that for any concave-down dispersion relation, such as \eqref{eq:disp} and \eqref{eq:disp-trunc}, the resonance condition in frequency $\omega$ cannot be satisfied if all wavenumbers involved share the same sign (i.e., all positive or all negative). Non-trivial resonant interactions occur only when one wavenumber has the opposite sign to the other three (e.g., one negative and three positive), or when two wavenumbers are positive and two are negative. The two-positive-two-negative configuration, however, requires at least one large wavenumber on the descending branch \eqref{eq:desc_branch}. Since both models have deviated substantially from the true shallow-water dispersion in this range, these interactions lie outside the regime of interest (i.e., $\kappa>\kappa_{\mathrm{max}}$), are therefore regarded as \textit{spurious} solutions that may be discarded. A graphical proof of the resonance configurations, exemplified using \eqref{eq:disp-trunc}, is provided in Appendix \ref{appB}, and is fundamental to the subsequent analysis of $T_{k123}$.

We begin our analysis with the KB equation with the dispersion relation \eqref{eq:disp}. According to the literature \citep[e.g.,][]{Zakharov1980,Zakharov1988,Zakharov1991}, we expect this system to yield a trivial interaction coefficient on the resonant manifold due to its complete integrability. Proving the coefficient vanishes on the resonant manifold for \eqref{eq:disp} is non-trivial, as the manifold lacks a simple analytical expression due to the square root. The next best approach is to verify numerically whether \eqref{Tkernek} vanishes on the resonant manifold. To this end, we compute $T_{k123}$ for a large number of resonant quartets using the dispersion relation \eqref{eq:disp}, with all computations carried out using quadruple precision. table \ref{tab:table1} presents a subset of these quartets tracing the resonant manifold, demonstrating that the resulting coefficient vanishes to numerical precision, which implies that the KB equation yields a trivial WKE.

\begin{table}
  \begin{center}
\def~{\hphantom{0}}
    \begin{tabular}{ccccc}
    \textrm{$k$} &
    \textrm{$k_1$} &
    \textrm{$k_2$} &
    \textrm{$k_3$} &
    \textrm{$|T_{k123}|$} \\[3pt]
$5.0$ & $1.982387$ & $7.0$  & $-1.76\times10^{-2}$ & $4.66\times10^{-32}$ \\
$5.0$ & $2.969726$ & $8.0$  & $-3.03\times10^{-2}$ & $4.47\times10^{-31}$ \\
$5.0$ & $3.954433$ & $9.0$  & $-4.56\times10^{-2}$ & $6.76\times10^{-31}$ \\
$5.0$ & $4.936444$ & $10.0$ & $-6.36\times10^{-2}$ & $3.50\times10^{-31}$ \\
$5.0$ & $5.915683$ & $11.0$ & $-8.43\times10^{-2}$ & $4.36\times10^{-31}$ \\
$5.0$ & $6.892058$ & $12.0$ & $-1.08\times10^{-1}$ & $7.05\times10^{-31}$ \\
$5.0$ & $7.865463$ & $13.0$ & $-1.35\times10^{-1}$ & $2.63\times10^{-31}$ \\
$5.0$ & $8.835773$ & $14.0$ & $-1.64\times10^{-1}$ & $4.30\times10^{-31}$ \\
$5.0$ & $9.802847$  & $15.0$ & $-1.97\times10^{-1}$ & $1.58\times10^{-31}$ \\
$5.0$ & $10.766521$ & $16.0$ & $-2.33\times10^{-1}$ & $1.86\times10^{-31}$\\
  \end{tabular}
\caption{\label{tab:table1}%
Values of the interaction coefficient $T_{k123}$ for the KB equation using $\beta=0.001$ for different resonant quartets.}
  \end{center}
\end{table}

Next, we move on to the analytical calculation for the non-integrable Boussinesq variant with dispersion relation \eqref{eq:disp-trunc}. The existing view for the shallow-water wave interaction coefficient, such as \eqref{Tkernek}, is that it contains both near-singular and regular terms \citep{Zakharov1999,Onorato2009}. The former correspond to frequency mismatches of $O(\beta)$ (i.e., their inverses are $O(1/\beta)$, which is near-singular), while the latter are of $O(1)$. It has previously been thought that the near-singular terms dominate the coefficient $T_{k123}$. However, the near-singular vs. regular terms need to be evaluated through a much more delicate analysis, which we perform below.  

Without loss of generality, we consider $k,k_1,k_2$ to be positive and $k_3$ to be negative, such that $\kappa_2 > \kappa, \kappa_1, \kappa_3$. Under this assumption, the ``near-singular'' and ``regular'' terms in \eqref{Tkernek} (hereafter named as $T_{\mathrm{I}}$ and $T_{\mathrm{II}}$) can be collected as 
\begin{subequations}\label{T12}
\begin{gather}
    T_{\mathrm{I}} =
    -\frac{V_{k,3,k-3}V_{2,1,2-1}}{\omega_1+\omega_{2-1}-\omega_2}
    -\frac{V_{1,3,1-3}V_{2,k,2-k}}{\omega_k+\omega_{2-k}-\omega_2}
    -\frac{V_{k+1,k,1}V_{2+3,2,3}}{\omega_{k+1}-\omega_k-\omega_1},
    \label{T12a}\\
    T_{\mathrm{II}} = T_{k123}-T_{\mathrm{I}}.
    \label{T12b}
\end{gather}
\end{subequations}
Substitution of \eqref{eq:disp-trunc} and \eqref{Vkernel} into \eqref{T12} gives (see details in Appendix \ref{appC})
\begin{subequations}\label{T12_res}
\begin{gather}
    T_{\mathrm{I}} = -\frac{3\alpha^2}{16\beta}
    \Big( \frac{\kappa_3^3}{\kappa\kappa_1\kappa_2} \Big)^{1/2},
    \label{T12_resa}\\
    T_{\mathrm{II}}= -\frac{\alpha^2}{64}
    (\kappa\kappa_1\kappa_2\kappa_3)^{1/2}
    \Big[\kappa_3\Big(\frac{1}{\kappa_1}-\frac{1}{\kappa_2}+\frac{1}{\kappa}\Big)\Big].
    \label{T12_resb}
\end{gather}
\end{subequations}

At first glance, the relative magnitude of these two contributions in the small-$\beta$ limit is not immediately clear from \eqref{T12_res}. However, solving the resonance condition in \eqref{WKE} gives $\kappa_3=(\beta/4) \kappa \kappa_1 \kappa_2$ (see Appendix \ref{appD}); substituting this into \eqref{T12_res} yields
\begin{subequations}\label{T12_finres}
\begin{gather}
    T_{\mathrm{I}} = -\frac{3\alpha^2}{128}\beta^{1/2}\kappa\kappa_1\kappa_2,
    \label{T12_finresa}\\
    T_{\mathrm{II}} = -\frac{\alpha^2}{512}\beta^{3/2}\kappa\kappa_1\kappa_2
    (\kappa\kappa_2+\kappa_1\kappa_2-\kappa\kappa_1).
    \label{T12_finresb}
\end{gather}
\end{subequations}
Thus, for small $\beta$, the $T_{\mathrm{I}}$ dominates, and the total coefficient can be reduced to the compact leading-order form
\begin{equation}
T_{k123}\approx T_{\mathrm{I}}=-\frac{3\alpha^2}{128}\beta^{1/2}\kappa\kappa_1\kappa_2.
    \label{T_finres}
\end{equation}

We note that if any wavenumber other than $k_3$ is chosen to be negative (with the others positive), the coefficient $T_{k123}$ can be directly evaluated using its symmetry properties. For example, if $k_1<0$, one can use the relation $T_{k123}=T_{231k}$, where the latter evaluates to $T_{231k}=-(3\alpha^2/128)[\sqrt{\beta}\kappa\kappa_2\kappa_3]$. Consequently, a general expression for $T_{k123}$ can be written as
\begin{equation}
    T_{k123} = -\frac{3\alpha^2}{128} \Big(\sqrt{\beta}\kappa_{\mathrm{mul}}\Big), 
    \label{Tk123_fin}
\end{equation}
where $\kappa_{\mathrm{mul}}$ represents the multiplication of the magnitudes of the three wavenumbers with the same sign.

We also remark here that it is claimed in \citet{Zakharov1999} that all ``near-singular'' terms in $T_{k123}$ cancel out, based on the formulation from the Euler equation. This claim is in fact not relevant to the present case, since \citep{Zakharov1999} assumes that all wavenumbers share the same sign, which is not consistent with the resonant manifold in \eqref{WKE}.

Since $T_{k123}\neq 0$ on the resonant manifold according to \eqref{Tk123_fin}, we have obtained a non-trivial WKE \eqref{WKE} from \eqref{eq:kb_n1} and \eqref{eq:kb_n2} with a truncated-dispersion relation approximation. Next, we seek stationary solutions to \eqref{WKE}. The first set of stationary solutions is of Rayleigh-Jeans type, namely the equipartition-of-action solution $n_k\sim \kappa^0$ and the equipartition-of-energy solution $n_k\sim 1/\omega_{k}\cong1/\kappa^{1}$. These naturally hold for \eqref{WKE} since they turn the integrand in the integral on the right-hand side into zero at each point. However, the Kolmogorov-Zakharov turbulent solution to \eqref{WKE} needs to be found in a novel way, which we describe in the following section.   
\subsection{Kolmogorov-Zakharov spectra}\label{KZ}
One might attempt to find the KZ solution $n_k \sim\kappa^\gamma$ to \eqref{WKE} using a ``standard'' approach using the formula $\gamma=-2\tau/3 - 1$ \citep{Nazarenko2011} with $\tau$ the degree of homogeneity of $T_{k123}$. However, this approach does not hold here because the dispersion relation \eqref{eq:disp-trunc} is inhomogeneous, which breaks the basis to derive the ``standard'' formula. As one may have noticed here, even the degree of homogeneity $\tau$ is not certain here, depending on whether \eqref{T12_res} or \eqref{T12_finres} is used for evaluation, which results from the same inhomogeneity issue. If one simply considers a dispersion relation $\omega_k=\kappa$ with $\beta=0$, then the resonance condition leads to $\kappa_3=0$ in \eqref{T12_res} so that $T_{k123}=0$ and the problem becomes trivial. Therefore, to obtain the KZ spectrum, we have to develop a refined approach which considers the cubic terms in the dispersion relation (as we have done in deriving \eqref{T12_finres}).

We start by considering an isotropic spectrum, $n_{-k}=n_k$, which allows us to convert wavenumber $k$ into its magnitude $\kappa$ in \eqref{WKE}. In particular, the resonance condition in $k$ permits eight sign choices for $k_1$, $k_2$ and $k_3$ (for fixed positive $k$), but disregarding trivial pairings (e.g., $k=k_2$, $k_1=k_3$; $k_1=k_2$, $k=k_3$) leaves only four non-trivial cases where the integrand is non-zero. The resulting equation in $\kappa$ is
\begin{equation}
\begin{split}
    \frac{\partial n_\kappa}{\partial t} =&\ 4\pi \int |T_{k 123}|^2 n_\kappa n_1n_2n_3\left[ \frac{1}{n_\kappa}+\frac{1}{n_1}-\frac{1}{n_2}-\frac{1}{n_3} \right]\delta(\omega_\kappa+\omega_1-\omega_2-\omega_3)\\
    &\times\big[ \delta(\kappa+\kappa_1-\kappa_2+\kappa_3)
    +\delta(\kappa+\kappa_1+\kappa_2-\kappa_3)+\delta(\kappa-\kappa_1-\kappa_2-\kappa_3)\\
    &+\delta(\kappa-\kappa_1+\kappa_2+\kappa_3)\big]  d\kappa_1d\kappa_2d\kappa_3.
    \label{WKE-nt} 
\end{split}
\end{equation}

We name the four terms in \eqref{WKE-nt} $I_1$, $I_2$, $I_3$, and $I_4$, corresponding to the four delta functions in the square brackets. To further simplify $I_1$ associated with $\delta(\kappa+\kappa_1-\kappa_2+\kappa_3)$ in \eqref{WKE-nt} (as an example), we notice that the frequency resonance condition can be reduced to (see Appendix \ref{I-reduct})
\begin{equation}
    0=\omega_\kappa+\omega_1-\omega_2-\omega_3\approx\kappa_3-\frac{\beta}{4}\kappa\kappa_1\kappa_2.
\end{equation}
The wavenumber resonance condition is reduced to $\kappa+\kappa_1-\kappa_2+O(\beta)=0$, with the $O(\beta)$ term negligible at leading order. Therefore, integrating out the frequency condition and using \eqref{Tk123_fin} reduces $I_1$ to
\begin{equation}
\begin{split}
I_1=&\int\Big|\frac{3\alpha^2}{128}\sqrt{\beta}\kappa\kappa_1\kappa_2\Big|^2 n_\kappa n_1n_2n_3\left[ \frac{1}{n_\kappa}+\frac{1}{n_1}-\frac{1}{n_2}-\frac{1}{n_3} \right]\delta(\kappa+\kappa_1-\kappa_2)d\kappa_1d\kappa_2.
\end{split}
\label{reduced-I1}
\end{equation}
Following similar procedures, we can reduce $I_2$ and $I_3$ to

\begin{equation}
\begin{split}
I_2=&\int\Big|\frac{3\alpha^2}{128}\sqrt{\beta}\kappa\kappa_1\kappa_3\Big|^2 n_\kappa n_1n_2n_3\big[ \frac{1}{n_\kappa}+\frac{1}{n_1}-\frac{1}{n_2}-\frac{1}{n_3} \big]\delta(\kappa+\kappa_1-\kappa_3)d\kappa_1d\kappa_3
\end{split}
\label{reduced-I2}
\end{equation}
and
\begin{equation}
\begin{split}
I_3=&\int\Big|\frac{3\alpha^2}{128}\sqrt{\beta}\kappa\kappa_2\kappa_3\Big|^2 n_\kappa n_1n_2n_3\left[ \frac{1}{n_\kappa}+\frac{1}{n_1}-\frac{1}{n_2}-\frac{1}{n_3} \right]\delta(\kappa-\kappa_2-\kappa_3)d\kappa_2d\kappa_3,
\label{reduced-I3}
\end{split}
\end{equation}
respectively.
For $I_4$, the frequency condition cannot be satisfied for $\kappa \sim O(1)$, thus $I_4=0$ and does not contribute to \eqref{WKE-nt}.

Integrals \eqref{reduced-I1}-\eqref{reduced-I3} can be further simplified depending on the size of $n_\kappa$ at $\kappa\sim \beta$ relative to $n_\kappa$ in the inertial range. In what follows, we discuss two cases: Cases A and B with $n_\kappa$ respectively small and large in the vicinity of $\kappa\sim \beta$. These two cases physically correspond to situations with and without large-scale damping in the simulations demonstrated in section \S\ref{force-dissip}.

\subsubsection{Case A}\label{caseA}
Consider $I_1$ with $n_3\ll n_\kappa,n_1,n_2$, the integral in \eqref{reduced-I1} can be reformulated as (with $1/n_3$ term dominating over other terms in the square bracket)
\begin{equation}
I_1=-\int\Big|\frac{3\alpha^2}{128}\sqrt{\beta}\kappa\kappa_1\kappa_2\Big|^2 n_\kappa n_1n_2\delta(\kappa+\kappa_1-\kappa_2)d\kappa_1d\kappa_2,
\label{I1-3w}
\end{equation}

Similar procedures applied to $I_2$ and $I_3$ reduce \eqref{reduced-I2} and \eqref{reduced-I3} to
\begin{equation}
I_2=-\int\Big|\frac{3\alpha^2}{128}\sqrt{\beta}\kappa\kappa_1\kappa_3\Big|^2 n_\kappa n_1n_3\delta(\kappa+\kappa_1-\kappa_3)d\kappa_1d\kappa_3,
\label{I2-3w}
\end{equation}
and
\begin{equation}
I_3=\int\Big|\frac{3\alpha^2}{128}\sqrt{\beta}\kappa\kappa_2\kappa_3\Big|^2 n_\kappa n_2n_3\delta(\kappa-\kappa_2-\kappa_3)d\kappa_2d\kappa_3.
\label{I3-3w}
\end{equation}

By defining 
\begin{equation}
R^\kappa_{12}=\Big|\frac{3\alpha^2}{128}\sqrt{\beta}\kappa\kappa_1\kappa_2\Big|^2n_\kappa n_1n_2\delta(\kappa-\kappa_2-\kappa_1)
\label{int-A}
\end{equation}
the integral $I$ can be expressed as
\begin{equation}
    I=I_1+I_2+I_3=\int(R^\kappa_{12}-R^2_{1\kappa}-R^1_{\kappa2})d\kappa_1d\kappa_2.
\label{total-I}
\end{equation}
After applying the Zakharov transformations \citep{Zakharov2012} and assuming a solution ansatz $n_\kappa \sim \kappa^{-x}$ (see Appendix \ref{caseA-app} for full details), we can write \eqref{total-I} as
\begin{equation}
I =\int R^\kappa_{12}\Big[1-\Big(\frac{\kappa_1}{\kappa}\Big)^{3x-8}-\Big(\frac{\kappa_2}{\kappa}\Big)^{3x-8}\Big]d\kappa_1d\kappa_2.
\label{total-I-A}
\end{equation}
The condition for a stationary solution is obtained by requiring the bracketed term to vanish, which leads to
\begin{equation}
3x-8=1 \;\Rightarrow\; x=3,
\label{KZ-A}
\end{equation}
which is the KZ exponent for Case A with small $n_\kappa$ at $\kappa\sim\beta$.

In Appendix \ref{app-local}, we further provide a locality analysis which leads to a locality window of $5/2<x<4$. The solution $x=3$ is therefore within the locality window and represents a valid local solution to the WKE.

\subsubsection{Case B}\label{caseB}
Considering $n_3\gg n_\kappa,n_1,n_2$ for $I_1$ and similar conditions for $I_2$ and $I_3$, integrals \eqref{reduced-I1}-\eqref{reduced-I3} are reduced to
\begin{equation}
\begin{split}
I_1=&\int\Big|\frac{3\alpha^2}{128}\sqrt{\beta}\kappa\kappa_1\kappa_2\Big|^2n_\kappa n_1n_2n_3 \big[ \frac{1}{n_\kappa}+\frac{1}{n_1}-\frac{1}{n_2} \big]\delta(\kappa+\kappa_1-\kappa_2)d\kappa_1d\kappa_2,
\end{split}
\label{I1-3wB}
\end{equation}

\begin{equation}
\begin{split}
I_2=&\int\Big|\frac{3\alpha^2}{128}\sqrt{\beta}\kappa\kappa_1\kappa_3\Big|^2 n_\kappa n_1n_2n_3\big[ \frac{1}{n_\kappa}+\frac{1}{n_1}-\frac{1}{n_3} \big]\delta(\kappa+\kappa_1-\kappa_3)d\kappa_1d\kappa_3
\end{split}
\label{I2-3B}
\end{equation}
and
\begin{equation}
\begin{split}
I_3=&\int\Big|\frac{3\alpha^2}{128}\sqrt{\beta}\kappa\kappa_2\kappa_3\Big|^2 n_\kappa n_1 n_2n_3\big[ \frac{1}{n_\kappa}-\frac{1}{n_2}-\frac{1}{n_3} \big]\delta(\kappa-\kappa_2-\kappa_3)d\kappa_2d\kappa_3.
\end{split}
\label{I3-3B}
\end{equation}
Defining
\begin{equation}
R^\kappa_{12}=\Big|\frac{3\alpha^2}{128}\sqrt{\beta}\kappa\kappa_1\kappa_2\Big|^2n_\kappa n_1n_2n_3\big[ \frac{1}{n_\kappa}-\frac{1}{n_2}-\frac{1}{n_1} \big]\delta(\kappa-\kappa_2-\kappa_1),
\label{int-B}
\end{equation}
and applying the Zakharov transformations with the solution ansatz $n_\kappa\sim \kappa^{-x}$ (see Appendix \ref{caseB-app} for full details), we arrive at
\begin{equation}
I =\int R^\kappa_{12}\Big[1-\Big(\frac{\kappa_1}{\kappa}\Big)^{6x-8}-\Big(\frac{\kappa_2}{\kappa}\Big)^{6x-8}\Big]d\kappa_1d\kappa_2.
\label{total-I-B}
\end{equation}
By enforcing the same condition as in \S\ref{caseA}, we find that the stationary solution is obtained at
\begin{equation}
6x-8=1 \;\Rightarrow\; x=\frac{3}{2},
\label{KZ-B-sol}
\end{equation}
which is the KZ exponent for Case B with large $n_\kappa$ at $\kappa\sim\beta$. The locality analysis in Appendix \ref{app-local} shows a locality window of $4/3<x<5/2$, so that this KZ solution is local.

 \section{Numerical study}
A subsequent numerical study is performed with the purpose of verifying the theoretical results derived in \S\ref{sec:theoretical-deriv}. In this section, we begin by outlining the methodology and numerical setup implemented in this study. We then discuss the results from two scenarios, namely, free-evolution and forced-dissipated configurations. 

\subsection{Methodology}
We simulate the KB system \eqref{eq:kb_n1} and \eqref{eq:kb_n2} using a pseudospectral method paired with a fourth-order Runge-Kutta time marching scheme with an integration factor (IF-RK4) formulation \citep[e.g.,][]{Pan2020}. Key details of the scheme can be found in Appendix \ref{appE}, along with a validation against existing exact solutions of the KB system. All numerical experiments are performed on a computational domain of size $L = 2\pi $ with periodic boundary conditions, discretised using $N = 4096$ free wave modes (before de-aliasing). The value of $\beta$ is correspondingly chosen (and specified in each subsection) such that the high-order term in the dispersion relation in \eqref{eq:disp} is much smaller than the linear term, even for the highest wavenumber. In this section, we further present details of the numerical setup for the free-evolution and forced-dissipated cases in \S\ref{free-evo} and \S\ref{force-dissip}, respectively. We then introduce the formulation of the direct scattering transform (DST), as well as the numerical strategies to solve the spectral problem for the purpose of testing isospectrality for the KB model associated with its integrability.

\subsubsection{Free-evolution configuration}\label{free-evo-config}
All freely evolving cases are performed without external forcing and damping, using a Gaussian spectrum $S(\kappa)$ as the initial condition \citep[e.g.,][]{Pelinovsky2006,Flamarion2024} 
\begin{equation}
    S(\kappa)=Q\exp\bigg(-\frac{(\kappa-\kappa_p)^2}{2K^2}\bigg),
\end{equation}
where $\kappa_p=512$ represents the wavenumber peak, $K=51.2$ characterises the spectral bandwidth, and $Q$ is chosen such that the resulting data size is $\lesssim 1$ (say measured by $||\hat{\eta}_\kappa||_\infty$ and $||\hat{u}_\kappa||_\infty$). The amplitude of each Fourier component $\hat{\eta}_\kappa$ is drawn from the spectrum $S(\kappa)$. To construct a bidirectional field, $\hat{\eta}_\kappa$ is evenly split to equally distribute energy between left- and right-propagating components. Each component is assigned a random phase drawn uniformly from $(0,2\pi]$. The component $\hat{u}_\kappa$ is then obtained from $\eta_\kappa$ for both left- and right-propagating components via the linear relation. For all free-evolution experiments, we fix the dispersion parameter $\beta=0.2$, and vary the nonlinearity parameter $\alpha$ to control the nonlinearity level.
\subsubsection{Forced-dissipation configuration}\label{force-dissip-config}
All experiments begin from a quiescent initial state. Waves are excited by a forcing term $F(\kappa,t)$ added to the right-hand side of \eqref{eq:kb_n2} (or \eqref{eq:kb_F2} defined in the spectral domain), in the form of \citep[e.g.,][]{Dyachenko2004,Pan2020,Zhang2022,Korotkevich2023}

\begin{equation}
F(\kappa,t)=
\begin{cases}
f_\kappa \exp\!\big[-C t + i(\omega_\kappa t + R)\big], & t \le T_c,\\[4pt]
f_\kappa \exp\!\big[-C T_c +i(\omega_\kappa t + R)\big], & t > T_c,
\end{cases}
\label{eq:forcing}
\end{equation}
with 
\begin{equation}
f_\kappa=
\begin{cases}
f_0 \frac{(\kappa-\kappa_1)(\kappa_2-\kappa)}{(\kappa_1-\kappa_2)^2}, & \kappa_1\le \kappa \le \kappa_2,\\[4pt]
0, & \textnormal{otherwise},
\end{cases}
\label{eq:force_amp}
\end{equation}
where $f_0=0.88$ controls the forcing amplitude (leading to a solution of size $O(1)$ and mitigating the finite-size effect \citep[e.g.,][]{Lvov2010,Hrabski2020,Zhang2022b}), $\kappa_1=26$ and $\kappa_2=38$ define the lower and upper bounds of the forcing band, and $R$ is a uniformly distributed random number in the interval $(0,2\pi]$. As defined in \eqref{eq:forcing}, the forcing amplitude decays exponentially at rate $C$ for $t<T_c$ and remains constant for $t\ge T_c$. Such an implementation allows for a stationary state of the system to be achieved faster \citep{Zhang2022}. For this study, we set $T_c=500T_p$ and $C=\mathrm{ln}5/T_c$, where $T_p$ is the peak period (computed using $\kappa_p=(\kappa_1+\kappa_2)/2=32$). 

To account for dissipation at small scales, we add terms $D_1\hat{\eta}_\kappa$ and $D_1\hat{u}_\kappa$ to \eqref{eq:kb_n1} and \eqref{eq:kb_n2}, respectively. The dissipation coefficient takes the form \citep[e.g.,][]{Dyachenko2004,Korotkevich2023}
\begin{equation}
D_1(\kappa)=\nu_1 (\kappa-\kappa_{d_1})^{2}, \qquad \kappa \ge \kappa_{d_1}
\end{equation}
where $\nu_1=10^{-4}$ controls the dissipation strength, and $\kappa_{d_1}=900$ is the onset wavenumber. To distinguish Cases A and B, we introduce a hypo-viscous damping term at large scales only for Case A, with the coefficient in the form of
\begin{equation}
    D_2(\kappa)=\nu_2 \kappa^{-4}, \qquad \kappa \le \kappa_{d_2}.
\label{large-damp}
\end{equation}
where $\nu_2=5\times10^{-6}$. When \eqref{large-damp} is applied, it introduces damping that suppresses $n_\kappa$ at large scales, consistent with the assumption in Case A. Without this damping, the inverse cascade fills in large scales, causing $n_k$ to grow large in the simulation, consistent with Case B. In all forced-dissipated cases, we fix $\alpha=0.0001$ and $\beta=0.001$, so that the validity condition of normal-form transformation is satisfied together with $O(1)$ solution size.

\subsubsection{Direct scattering transform}
In this study, we utilise the direct scattering transform (DST) as a diagnostic tool. Since the KB system is completely integrable, it admits a Lax pair. We focus on the spectral problem alone, which suffices for the solution of DST. At each time instant, the spectral problem for the system is given by \citep{Kaup1975}
\begin{equation}
\Psi_{xx}+\big[k^2+\frac{3}{4}\beta^{-1}+ikq(x)+r(x)\big]\Psi=0
\label{spec-k}
\end{equation}
where $k$ is the eigenvalue, $\Psi$ is the eigenfunction, and $q(x)$ and $r(x)$ are potentials defined as
\[
q(x)=\frac{\sqrt3}{2}\alpha\beta^{-1/2}u(x), \quad r(x)=\frac{3}{4}\alpha\beta^{-1}\big(\eta(x)-\frac{1}{4}\alpha u(x)^2\big),
\]
which are assumed to decay to zero at $|x|\rightarrow \infty$. Solving the DST problem \eqref{spec-k} amounts to finding the eigenvalue $k$ and its corresponding eigenfunction $\Psi$. The property of the solution depends critically on the spectral parameter defined by $E(k)=\{k^2+3/(4\beta)\}^{1/2}$. However, $E(k)$ is a multivalued function in the complex plane of $k$, and involves branch points at $k=\pm i\sqrt{3/(4\beta)}$ (more precisely, when a closed loop around $\pm i\sqrt{3/(4\beta)}$ is followed, $E(k)$ changes its value as the loop returns to the original point).  
Since it is inconvenient to work with a branch point, it is a conventional practice to perform a conformal transformation to the $\zeta$-plane to eliminate these branch points \citep{Kaup1975}. The transformation is defined by 
\[
k=\frac{1}{4}\big(\zeta-\frac{B}{\zeta}\big), \qquad E=\frac{1}{4}\big(\zeta+\frac{B}{\zeta}\big)
\]
with $B = 3/\beta$ so that $E(\zeta)=\{k(\zeta)^2+3/(4\beta)\}^{1/2}$ as defined above. The spectral problem \eqref{spec-k} can now be written as
\begin{equation}
\Psi_{xx}+\big[E(\zeta)^2+ik(\zeta)q(x)+r(x)\big]\Psi=0,
\label{spec-zeta}
\end{equation}
where both $k$ and $E$ are single-valued in $\zeta$. We have now circumvented the multivalued function and branch points that are associated with working in the $k$-plane. As a result, the transformation $\zeta\mapsto-B/\zeta$ changes the sign of $E$, $E\mapsto-E$, and leaves both $k$ and $E^2$ invariant. Therefore, each $\zeta$ solution of \eqref{spec-zeta} has a ``mirror" part of $-B/\zeta$, and we plot in the result section only the one corresponding to the upper half $E$-plane, i.e., with $\mathrm{Im}[E(\zeta)]>0$ (see caption of figure \ref{fig:case_B} for details).

Referring to \eqref{spec-zeta} again, we see that a continuous spectrum is characterised by $\mathrm{Im}[E(\zeta)]=0$, since at $|x|\rightarrow \infty$, it corresponds to an oscillatory solution of $\Psi(x)$. This solution occurs when $\zeta$ is either purely real or lies on the curve defined by $|\zeta|=\sqrt{3/\beta}$, as described in the caption of figure \ref{fig:disc_spec}. We are interested in the discrete spectrum of eigenvalues that are away from the continuous spectrum, physically corresponding to bound states (or solitons). 

The detailed numerical procedure to solve \eqref{spec-zeta} is described in Appendix \ref{appG}. We further remark that the traditional DST procedure assumes the problem to be defined on $\mathbb{R}$, but our numerical solutions of $u(x)$ and $\eta(x)$ are periodic. For this reason, the DST may not be precise; nevertheless, it remains an effective tool for identifying bound states of the solution (see \citet{Colleaux2025} for similar applications). In practice, this problem requires us to handle some spurious eigenvalues using a specific criterion detailed in Appendix \ref{appG}.



\subsection{Results}
\subsubsection{Free-evolution experiments}\label{free-evo}
\begin{figure}
\centerline{\includegraphics[]{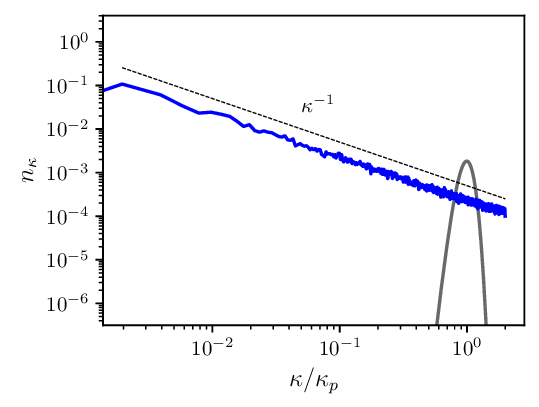}}
\caption{Initial (solid grey) and stationary (solid blue) spectra at $t=0$ and $t=2\times10^6T_p$ for the free-evolution case with $\alpha=\beta=0.2$. The thermal-equilibrium scaling is denoted (dashed black line).}
\label{fig:nk_zeta_spec}
\end{figure}

\begin{figure}
\centerline{\includegraphics[]{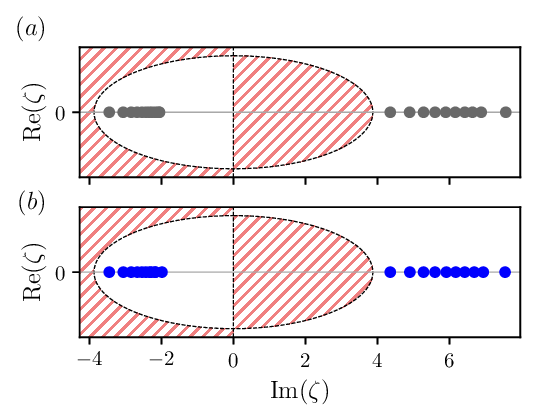}}
\caption{The spectrum of discrete eigenvalues $\zeta$ (blue and grey points) obtained via DST at $(a)$ $t=0$ and $(b)$ $t\approx2\times10^6 T_p$, corresponding to the times of initial and stationary spectra in figure\ref{fig:nk_zeta_spec}. The lower half $E$-plane (i.e., where $\textnormal{Im}[E(\zeta)]<0$) is marked by the red hashed region and the continuous spectrum region (i.e., where $\textnormal{Im}[E(\zeta)]=0$) is marked by the dashed black line.}
\label{fig:disc_spec}
\end{figure}

We start by examining a free-evolution case for the full dispersion relation \eqref{eq:disp} with $\alpha=0.2$ (and $\beta=0.2$ for all cases in this section as presented in \S\ref{free-evo-config}), which corresponds to a traditional regime with dispersion balancing nonlinearity. Figure \ref{fig:nk_zeta_spec} shows the evolution of the system from an initial Gaussian spectrum to the stationary spectrum evaluated at $t=2\times10^6T_p$. It is clear that the wavenumber spectrum exhibits significant evolution, ultimately reaching a thermal-equilibrium state characterised by $n_\kappa\sim \kappa^{-1}$. Figure \ref{fig:disc_spec} presents the corresponding discrete eigenvalue spectrum obtained via the DST at the beginning and end of the simulation (we note that the identification of bound states in a random wave background of high nonlinearity is also found in \citet{Colleaux2025}). Aside from some very minor drift attributed to the periodic domain (in contrast to the unbounded domain $\mathbb{R}$ assumed in theory), all eigenvalues remain essentially constant throughout the long-time evolution, consistent with the isospectral property of the integrable system. This case therefore provides clear evidence that complete integrability does not guarantee a frozen wavenumber spectrum, as substantial spectral evolution can still occur.

\begin{figure}
\centerline{\includegraphics{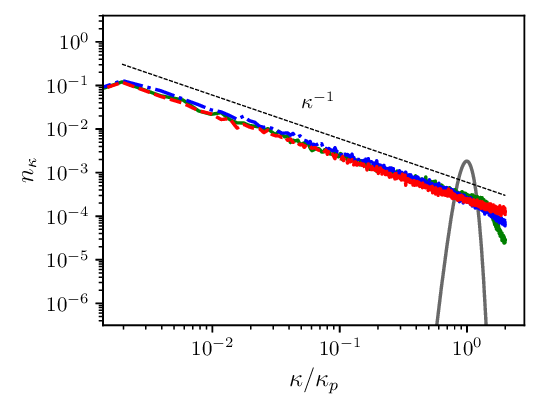}}
\caption{The initial spectrum (solid grey) and stationary spectra with $\alpha=0.2$ (dashed red), $\alpha=0.04$ (dash-dot blue) and $\alpha=0.02$ (solid green), respectively taken at $5\times10^5T_p$, $8\times10^5T_p$, and $1.2\times10^6T_p$. The thermal-equilibrium scaling is denoted (dashed black).}
\label{fig:free_decay_spec}

\end{figure}

\begin{figure}
  \centerline{\includegraphics{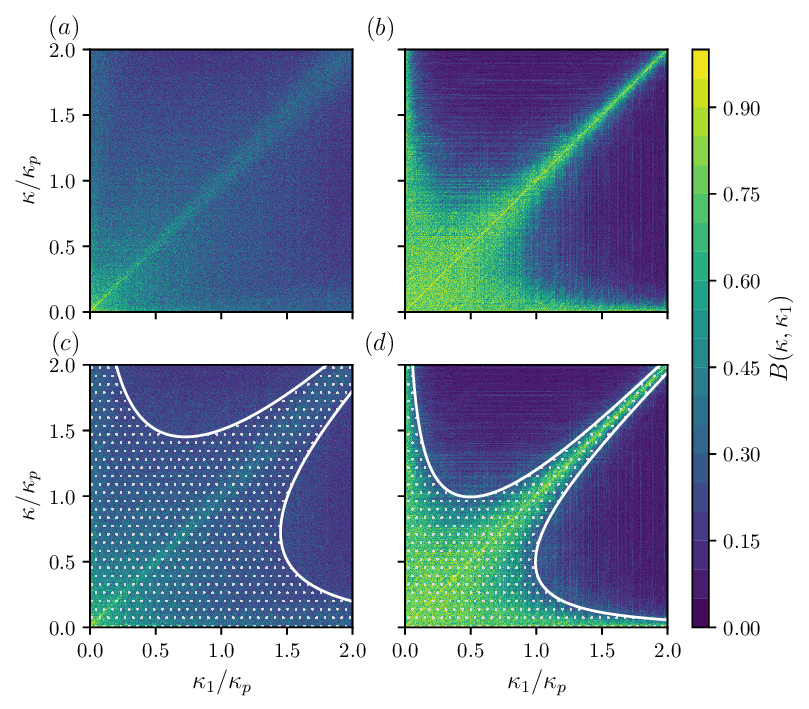}}
    \caption{Bi-coherence $B(\kappa,\kappa_1)$ for $(a,c)$ $\alpha=0.2$ and $(b,d)$ $\alpha=0.02$. Panels $(a)$ and $(b)$ show the full bi-coherence distribution, while panels $(c)$ and $(d)$ highlight the corresponding quasi-resonant regions. The solid white line marks the frequency-mismatch threshold $\Delta\omega=0.08$ for $\alpha=0.2$ and $\Delta\omega=0.025$ for $\alpha=0.02$, chosen such that the region enclosed (white dots) accounts for 80\% of the total bi-coherence.}
    \label{fig:bicoherence}
\end{figure}

\begin{figure}
  \centerline{\includegraphics{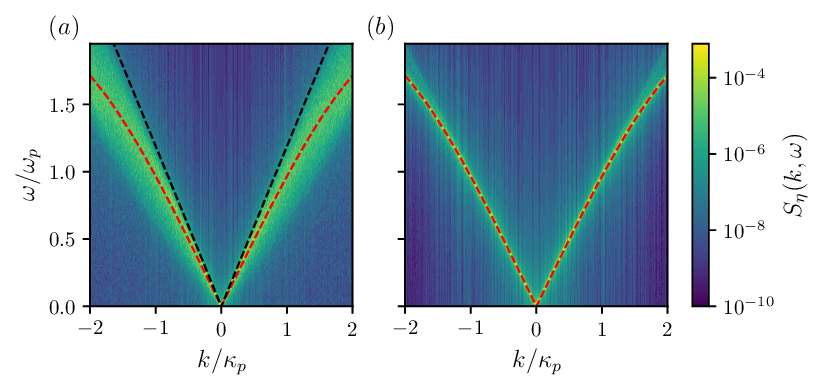}}
    \caption{The $k$-$\omega$ spectra $S_{\eta}(k,\omega)$ for $(a)$ $\alpha=0.2$ and $(b)$ $\alpha=0.02$. The KB dispersion relation \eqref{eq:disp} is denoted by the dashed red line and the soliton dispersion relation (for the most dominant bound state) is denoted by the dashed black line in $(a)$.}
    \label{fig:kw}
\end{figure}

We next study the spectral evolution for several values of $\alpha$, in order to cover a broad range of nonlinearity. Figure \ref{fig:free_decay_spec} shows the spectral evolution for $\alpha=0.2$, $0.04$ and $0.02$. We see that for all cases, the spectrum evolves into the same thermal-equilibrium state, although the corresponding evolution time differs (specifically, spectra are shown at $5\times10^5T_p$, $8\times10^5T_p$, and $1.2\times10^6T_p$ for the three cases). This spectral phenomenon has also been observed in numerical simulations of the KdV equation \citep{Pelinovsky2006}. In both cases, we suspect it arises from quasi-resonant three-wave interactions. To assess whether such interactions are relevant, we evaluate the bi-coherence defined as \citep{Pan2017}
\begin{equation}
    B(\kappa,\kappa_1)=\frac{|\langle\eta_{\kappa}^*\eta_{\kappa_1}\eta_{\kappa_2=\kappa-\kappa_1}\rangle|}{\langle|\eta_\kappa||\eta_{\kappa_1}||\eta_{\kappa_2=\kappa-\kappa_1}|\rangle}
    \label{B}
\end{equation}
where $\eta_\kappa$ is the Fourier component of the surface elevation obtained from numerical simulation and $\langle\cdots\rangle$ denotes a time average over the stationary data. The bi-coherence function $B(\kappa,\kappa_1)$ varies between 0 and 1, with 1 and 0 corresponding to perfect and no correlation, respectively. Given a set of discrete wavenumbers, $B$ obtains a high value for quasi-resonances. We compute $B(\kappa,\kappa_1)$ over all $(\kappa,\kappa_1)$ drawn from the discrete numerical grid. Figures \ref{fig:bicoherence}$(a,b)$ show the bi-coherence plots for high ($\alpha=0.2$) and low ($\alpha=0.02$) nonlinearity cases, respectively. In both cases, the analysis shows values of $B$ that are well above zero, confirming quasi-resonant triads across nonlinearity regimes. To robustly assess the nonlinearity dependence, we compute the frequency mismatch $\Delta\omega$ 
\begin{equation}
    |\omega(\kappa)-\omega(\kappa_1)-\omega(\kappa_2)|\le\Delta\omega
    \label{mismatch}
\end{equation}
for each discrete triad. For each case, we then determine the value of $\Delta\omega$ for which the cumulative bi-coherence of the triads satisfying \eqref{mismatch} accounts for 80\% of the total bi-coherence, and mark this threshold directly on the corresponding plot. Figure \ref{fig:bicoherence}$(c)$ demonstrates that the high nonlinearity enables quasi-resonant interactions across larger mismatches, as the allowable frequency mismatch $\Delta\omega$ grows with nonlinear broadening (evident in figure \ref{fig:kw}$(a)$). In contrast, the low nonlinearity case shown in figure \ref{fig:bicoherence}$(d)$ confines the quasi-resonances to a smaller frequency mismatch $\Delta\omega$. This confirms that quasi-resonant triads, active across all $\alpha$ regimes considered, drive the spectral evolution observed in this section.

We further remark that the spectral evolution towards $n_\kappa\sim \kappa^{-1}$ is also observed in experiments of bidirectional shallow-water waves, but it is often attributed to the soliton dynamics \citep[e.g.,][]{Redor2019,Redor2021,Leduque2025}. To assess the influence of solitons in our study, we plot the wavenumber-frequency ($k$-$\omega$) spectra from two representative simulations corresponding to the largest and smallest values of $\alpha$ in figure \ref{fig:kw}. We find that the energy branch in the $k$-$\omega$ spectra closely follows the free-wave dispersion relation \eqref{eq:disp}, rather than the soliton dispersion relation marked in the figure. For the case with the largest $\alpha=0.2$, the soliton energy identified by the DST can be computed and found to only constitute 3\% of the total energy, which is masked by the nonlinear broadening in figure \ref{fig:kw}$(a)$.  It is therefore safe to conclude that the thermalisation observed in our study is caused by interactions among random free waves, rather than by soliton dynamics.

\subsubsection{Forced-dissipated experiments}\label{force-dissip}
\begin{figure}
\centerline{\includegraphics[]{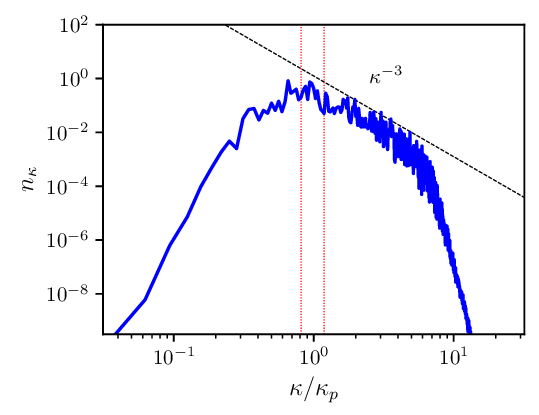}}
\caption{Stationary wave action spectrum for Case A. The forcing range is marked by the dotted red lines and the theoretical KZ scaling  $n_\kappa\sim \kappa^{-3}$ is denoted by the dashed black line.}
\label{fig:case_A}
\end{figure}

\begin{figure}
\centerline{\includegraphics[]{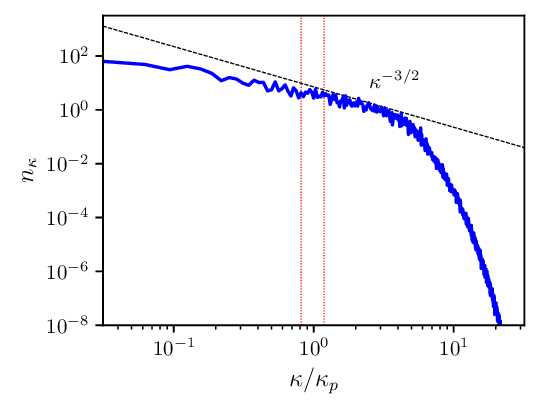}}
\caption{Stationary wave action spectrum for Case B. The forcing range is marked by the dotted red lines and the theoretical KZ scaling  $n_\kappa\sim \kappa^{-3/2}$ is denoted by the dashed black line.}
\label{fig:case_B}

\end{figure}

We now investigate the forced dissipated cases of the non-integrable Boussinesq equation. In this section, we use $\alpha=0.0001$ and $\beta=0.001$, ensuring that we operate in a regime where the normal-form transformation is valid and quartet interactions play a significant role in the spectral evolution towards small scales (see further discussion in Appendix \ref{appF}). We first present results with the presence of large-scale damping, corresponding to Case A in the theoretical derivation. Figure \ref{fig:case_A} shows the stationary spectrum after an evolution time of $6\times10^5T_p$, where large scales contain minimal energy as desired. We see an inertial range slightly exceeding 2/3 of a decade with a slope close to the theoretical prediction of $\kappa^{-3}$. Without large-scale damping, the resulting stationary spectrum corresponds to Case B with sufficiently high energy at large scales, as shown in figure \ref{fig:case_B}. To the right of the forcing range, we see the formation of a $\kappa^{-3/2}$ spectrum spanning approximately 2/3 of a decade, consistent with the theoretical prediction for Case B.

We remark that the results presented in this section differ substantially from the forced bidirectional shallow-water wave experiments in \citet{Redor2019,Redor2021,Leduque2025}. In these experiments, the system nonlinearity, measured by $\alpha/\beta$, is high enough to induce the formation of a large number of solitons (i.e., soliton gas), which manifests itself as an exponentially decaying energy spectrum at moderate to small scales. In our study, $\alpha/\beta$ is kept low, suppressing the formation of solitons. This is confirmed by DST showing no bound state content in the system. Therefore, the observed power-law KZ scalings arise solely from interactions of free random waves.

\section{Discussion and Conclusion}\label{discussion}

In this work, we establish the wave turbulence theory for 1-D bidirectional shallow-water waves derived from Boussinesq-type equations. We show that for the full KB equation, the four-wave interaction coefficient vanishes on the resonant manifold, as expected from integrability. However, when considering a truncated dispersion relation representative of non-integrable Boussinesq models, we are able to derive a robust, non-trivial wave kinetic equation as well as its KZ solutions.

Numerical simulations in this work highlight the significant spectral evolution that can be triggered by either quasi-resonant triad interactions or truncating the dispersion relation, which in turn breaks the integrability of the system. A closer examination of our results reveals that quasi-resonant quartet interactions may play a role in driving the kinetic behaviour, even in the full KB model. This may seem counterintuitive as it contradicts the empirical belief that integrable models do not yield kinetic behaviour. In the following, we briefly present two pieces of evidence, followed by a discussion of the mechanism.

The first piece of evidence comes from a set of free-evolution simulations of the full KB equation. In Appendix \ref{appF}, we provide identical simulations of the KdV equations, derived from KB by eliminating one direction of propagation (thus prohibiting four-wave interactions). For larger values of $\alpha$ (i.e., $\alpha=0.2$), the KdV simulation exhibits thermalisation on the same time scale as the KB simulation, which can be attributed to the quasi-resonant triad interactions. However, for smaller values of $\alpha$ (i.e., $\alpha=0.02$), only the KB simulation thermalises towards small scales. This illustrates that quasi-resonant quartet interactions must play a role, since exact resonances vanish on the resonant manifold.

The second piece of evidence comes from the forced-dissipated simulations. We find that even for the full KB equation, the spectrum evolves into a power-law state that appears to be consistent with both cases of the truncated-dispersion model (not shown due to close resemblance to figure \ref{fig:case_A} and figure \ref{fig:case_B}). This result is curious, as integrable models are not expected to exhibit significant spectral evolution due to the absence of wave kinetics. Since the values of $\alpha$ and $\beta$ are chosen to correspond to a regime where four-wave interactions are relevant, quasi-resonant quartet interactions likely play a significant role here.

\begin{figure}
\centerline{\includegraphics[]{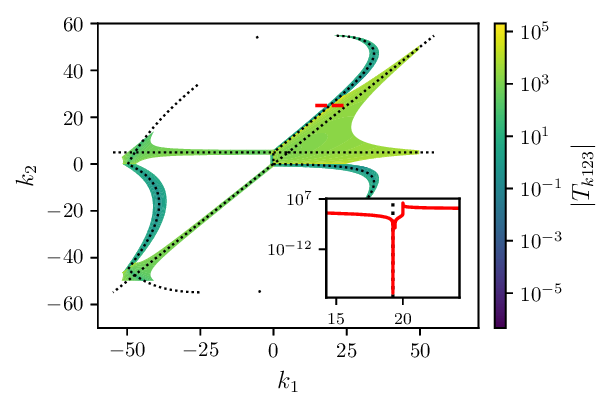}}
\caption{Interaction coefficient $T_{k123}$ for $\beta=0.001$, $k=5.0$, restricted to $|\Delta\omega|\le2$ near the KB resonant manifold. The dotted black lines mark $\Delta\omega=0$ curves: the curved branch is the KB resonant manifold, while the diagonal and horizontal branches correspond to trivial resonances. The inset shows a cross-section taken along the dashed red line, illustrating the behaviour of the coefficient across the resonant manifold.}
\label{fig:reson_manifold}
\end{figure}

\begin{figure}
\centerline{\includegraphics[]{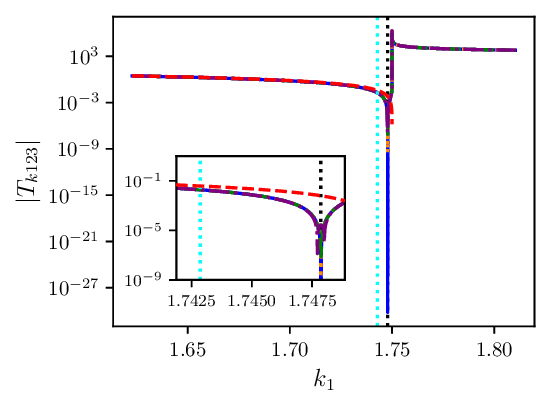}}
\caption{Interaction coefficients $T_{k123}$ for $\beta=0.001$, $k=1.5$, and $k_3=3.25$, evaluated using the KB dispersion relation (solid blue) and truncated dispersion relations with $O(\beta^3)$ corrections (dotted orange), $O(\beta^2)$ corrections (dashed green), and $O(\beta)$ corrections (dash-dotted purple). Also shown are the analytical coefficient \eqref{T12_resa} (dashed red), the KB resonant manifold (dotted black), and the contour $|\Delta\omega|=0.01$ (dotted cyan). The inset magnifies the behaviour near the resonant manifold.}
\label{fig:comp_kernel}
\end{figure}

To better understand the mechanism, we examine the behaviour of $T_{k123}$ near the resonant manifold, as shown in figure \ref{fig:reson_manifold}. In particular, the inset of the figure shows $T_{k123}$ along a section across the resonant manifold. We observe ``near-singular" behaviour in the sense that the coefficient drops sharply to zero on the resonant manifold, but quickly rises to significant finite values nearby. This implies that in the discrete setting with finite nonlinearity, even a minimal shift from the resonant manifold may allow quasi-resonant interactions to emerge as the primary mechanism for spectral evolution, while exact resonances may play a negligible role.
We provide further insight in figure \ref{fig:comp_kernel} by comparing $T_{k123}$ evaluated using the full KB dispersion relation with evaluations from truncated-dispersion approximations retaining successively higher-order terms. The figure also includes the analytical expression given by $T_\mathrm{I}$ in \eqref{T12_resa}. We see that $T_{k123}$ exhibits sharp minima on the resonant manifold, with each successive truncated-dispersion evaluation lying closer to the full KB result as higher-order dispersive corrections are retained. However, slightly off the resonant manifold ($|\Delta\omega|=0.01$), all coefficients are well described by the analytical expression. This small but finite frequency mismatch regime is most relevant for our numerical simulations (both full and truncated), since the discrete grid samples interactions near- rather than exact-resonance interactions. It is therefore unsurprising that the KB and non-integrable Boussinesq models exhibit the same KZ state, since in the vicinity of the resonant manifold, both coefficients coincide with the analytical expression. The dominant role of quasi-resonant interactions is consistent with examples from previous numerical studies of the Zakharov equation \citep{Annenkov2006} and MMT equation \citep{Hrabski2020}.

This work challenges the traditional view that there is no wave turbulence in the shallow-water setting. Quasi-resonances drive the kinetic behaviour in both the integrable KB and non-integrable models, aligning the former more closely with the traditional wave-turbulence framework and motivating further study of other integrable systems. A theoretical formulation of this kinetic behaviour and its analytical derivation is left for future work.

\backsection[Acknowledgements]{The authors thank Professor M. Onorato for his valuable insights and feedback on the interaction coefficient calculation, which substantially strengthened this work. In particular, he provided notes and code for the calculation of the KB coefficient, which helped us identify and correct an error in an early version of the manuscript.}

\backsection[Funding]{This research was supported by the Simons Foundation (Award ID \#651459). The computation was performed on the Great Lakes HPC Cluster provided by Advanced Research Computing (ARC) at the University of Michigan, Ann Arbor.}

\backsection[Declaration of interests]{The authors report no conflict of interest.}

\backsection[Author ORCIDs]{A. Simonis, https://orcid.org/0009-0000-3876-8160; S. Nazarenko, https://orcid.org/0000-0002-8614-4907; J. Shatah, https://orcid.org/0000-0003-3518-8069; Y. Pan, https://orcid.org/0000-0002-7504-8645}

\clearpage
\FloatBarrier
\appendix
\section{Validity condition for the normal-form transformation}
\label{appA}
\begin{figure}
\centerline{\includegraphics{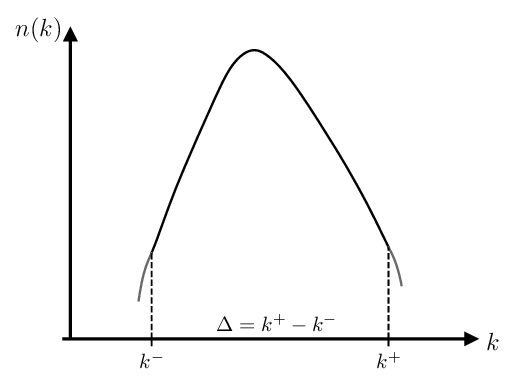}}
\caption{A sketch of the spectrum under consideration, with finite bandwidth $\Delta$.}
\label{fig:finspec}
\end{figure}

We consider a spectrum (see figure \ref{fig:finspec}) with finite bandwidth $\Delta=k^+ - k^-$, containing $N_\Delta = \Delta L/(2\pi)$ modes in $\Delta$. We make the following assumptions:
\begin{enumerate}
\item the spectrum decays sufficiently fast outside $[k^- ,k^+]$;
\item everything is comparable in the range $[k^- ,k^+]$, i.e., $k^+ = C_1 k^-$, $\mathrm{max}(n(k))=C_2 \mathrm{min}(n(k))$ for some constants $C_1$ and $C_2$.
\end{enumerate}
We note that assumption 1 is necessary to eliminate singularities arising from large wavenumbers in the normal-form transformation, as elaborated later in this appendix.
Our objective is to derive a condition under which \eqref{nearI} holds for $k\in p\mathbb{Z}$ where $p=2\pi/L$. For this purpose, let us consider the most singular terms on the left-hand side, i.e., those for which the denominators are of $O(\beta)$ rather than $O(1)$. For example, we take three wavenumbers in $\Gamma^{[1]}$ to be all positive, with $\omega_k-\omega_1-\omega_2 \sim \beta kk_1k_2$ in the first term and $\omega_k+\omega_1-\omega_2 \sim \beta kk_1k_2$ in the second term. Additionally, since $\omega_k+\omega_1+\omega_2 \sim O(1)$ always holds, the third term can be neglected. Putting everything together, we have

\begin{equation}
\begin{split}
     \mathrm{LHS} \sim& \Big|\sum_{k_1+k_2=k, \text{all} >0} \frac{\alpha (kk_1k_2)^{1/2}}{\beta kk_1k_2} b_{k_1}b_{k_2} -  2\sum_{k_1+k=k_2, \text{all} >0}  \frac{\alpha (kk_1k_2)^{1/2}}{\beta kk_1k_2} b_{k_1}^*b_{k_2} \Big|\\
        \sim&   \Big|\sum_{k_1=p}^{k-p}\frac{\alpha}{\beta [kk_1(k-k_1)]^{1/2}} b_{k_1}b_{k-k_1} + \sum_{k_1=p}^{\infty}\frac{\alpha}{\beta [kk_1(k+k_1)]^{1/2}} b_{k_1}^*b_{k+k_1} \Big| \\
        \sim&  \Big| \underbrace{\sum_{k_1=p}^{k-p}\frac{\alpha}{\beta [kk_1(k-k_1)]^{1/2}} b_{k_1}b_{k-k_1}}_{\circled{1}} +  \underbrace{\sum_{k_1=p}^{k-p}\frac{\alpha}{\beta [kk_1(k+k_1)]^{1/2}} b_{k_1}^*b_{k+k_1}}_{\circled{2}}\\
        &\ \ +  \underbrace{\sum_{k_1=k}^{\infty}\frac{\alpha}{\beta [kk_1(k+k_1)]^{1/2}} b_{k_1}^*b_{k+k_1}}_{\circled{3}} \Big|.
\end{split}
\label{LHSsim}
\end{equation}
In the above equation, it is evident that $\circled{1} > \circled{2}$ and 
\[
\begin{split}
   \circled{3} &< \sum_{k_1=k+p}^{\infty}\frac{\alpha}{\beta [kk_1|k-k_1|]^{1/2}} b_{k_1}^*b_{k+k_1}+ \frac{\alpha}{\beta [k^2(k+k)]^{1/2}} b_{k}^*b_{2k}\\
   &\sim \sum_{k_1=k+p}^{\infty}\frac{\alpha}{\beta [kk_1|k-k_1|]^{1/2}} b_{k_1}^*b_{k+k_1}.  
\end{split}   
\]
Finally, we have
\begin{equation}
   \mathrm{LHS} \sim \Big| \sum_{k_1=p, k_1\neq k}^{\infty}\frac{\alpha}{\beta [kk_1|k-k_1|]^{1/2}} b_{k_1}b_{k-k_1} \Big|.  
   \label{LHS_fin}
\end{equation}
Here, we note that if $b_k\sim O(1)$ for all $k$, the summation in \eqref{LHS_fin} (or \eqref{LHSsim}) does not converge at the $\infty$ end, as can be seen by replacing the summation by an integral. This problem is avoided with assumption 1 by introducing a finite spectral bandwidth $\Delta$. If one attempts to perform a more sophisticated analysis, the approach presented in \citet{Wu2025} can be followed.

In further evaluating \eqref{LHS_fin}, it is critical to take into account the random phases in $b_k$ (a procedure not considered in \citet{Zakharov1999}). More precisely, with random phases in $b_k$ the summation on the left-hand side behaves like a random walk in the complex plane, for which cancellations due to walks in different directions must be considered. For this reason, the summation on the left-hand side should be evaluated in terms of the expected value $E[|\sum \boldsymbol \cdot|]$ or $\{ E[|\sum \boldsymbol \cdot|^2] \}^{1/2}$. Taking the latter expression, we have
\begin{equation}
\begin{split}
   \mathrm{LHS} & \sim \frac{\alpha}{\beta k^{1/2}} \Bigg\{ E\bigg[ \Big| \sum_{k_1=p, k_1\neq k}^{\infty}\frac{b_{k_1}b_{k-k_1}}{ [k_1|k-k_1|]^{1/2}} \Big|^2 \bigg] \Bigg\}^{1/2} \\
       & \sim \frac{\alpha}{\beta k^{1/2}} \Bigg\{ \sum_{k_1=p, k_1\neq k}^{\infty} \frac{b_{k_1}b_{k_1}^*b_{k-k_1}b_{k-k_1}^*}{ [k_1|k-k_1|]} \Bigg\}^{1/2} \\
       & \sim \frac{\alpha |b_k|^2}{\beta k^{3/2}} \sqrt{N_{\Delta}}  \ll |b_k|,
\end{split}
\label{LHS_fin2}
\end{equation}
where we have used Wick's pairing to obtain the second line, and assumptions 1 and 2 to obtain the third line.

We now estimate $b_k$ using \eqref{eq:can_a} and \eqref{akbk}, which yields
\begin{equation}
   |b_k| \sim |\eta_k|/k^{1/2} \sim \frac{1}{\sqrt{N_\Delta}k^{1/2}},
   \label{bkest}
\end{equation}
where we have used $\sqrt{\sum_k |\eta_k|^2} \sim 1$ since the problem (e.g., significant wave height) has been scaled by $a$ in \eqref{eq:nond}. 

Combining \eqref{LHS_fin2} and \eqref{bkest}, we obtain $\alpha \ll \beta k^2$. If we further consider $k\sim 1$ (since the problem has been scaled by the peak wavenumber $k_p$), the condition is simply $\alpha \ll \beta$. Finally, we remark that the absence of $N_\Delta$ from the final result is only possible with the consideration of random phases in $b_k$.

\section{Graphical proof regarding the resonant conditions}
\label{appB}
\begin{figure}
\centerline{\includegraphics[]{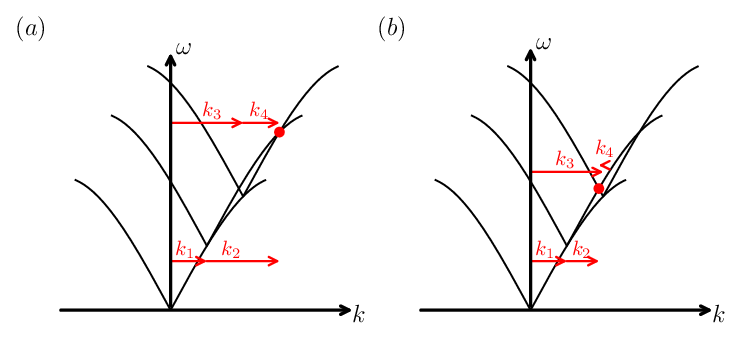}}
\caption{Graphical illustration of the resonant condition: $(a)$ the case where all wavenumbers are positive; $(b)$ the case where one wavenumber (say $k_4$) is negative and the others are positive.}
\label{fig:res_cond}
\end{figure}

\begin{figure}
\centerline{\includegraphics[]{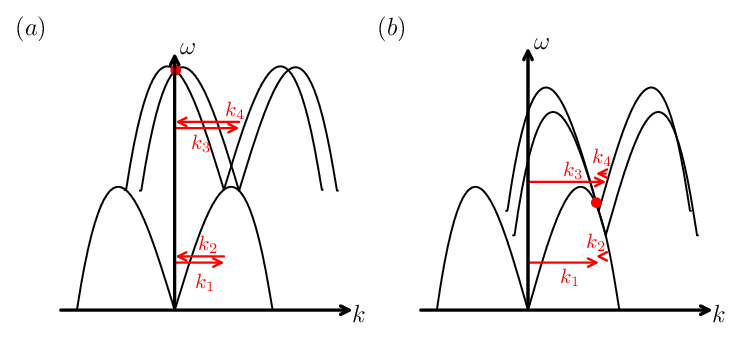}}
\caption{Graphical illustration of the spurious resonant conditions: $(a)$ the case where all wavenumbers are $O(\beta^{-1/2})$; $(b)$ the case where two wavenumbers are $O(1)$ and the other two are $O(\beta^{-1/2})$.}
\label{fig:res_cond-spur}
\end{figure}

We consider the resonant condition
\begin{equation}
   k_1+k_2=k_3+k_4,
\end{equation}
\begin{equation}
   \omega_1+\omega_2=\omega_3+\omega_4,
\end{equation}
with a concave-down dispersion relation such as \eqref{eq:disp-trunc} and \eqref{eq:disp}. For the sake of brevity, we will use the non-integrable Boussinesq equation with dispersion relation \eqref{eq:disp-trunc} as an example; however, the configurations generally hold for the full KB equation.  A graphical illustration of the resonant conditions is shown in figure \ref{fig:res_cond} and figure \ref{fig:res_cond-spur}. Figure \ref{fig:res_cond}$(a)$ represents the solution with all wavenumbers positive. However, in this case, this solution corresponds to the trivial solution of $k_1=k_4$ and $k_2=k_3$, which can be removed by a frequency renormalisation in the four-wave dynamical equation. Figure \ref{fig:res_cond}$(b)$ corresponds to the solution with one wavenumber (say $k_4$) negative and the other three positive, which persists in the WKE. Figures \ref{fig:res_cond-spur}$(a)$ and $(b)$ correspond to configurations with two positive and two negative wavenumbers. Such configurations require either (i) $|k_1|, |k_2|, |k_3|, |k_4| = O(\beta^{-1/2})$, as shown in figure \ref{fig:res_cond-spur}$(a)$, or (ii) $|k_2|, |k_4| = O(1)$ and $|k_1|, |k_3| = O(\beta^{-1/2})$, as shown in figure \ref{fig:res_cond-spur}$(b)$. In either case, at least one of the $O(\beta^{-1/2})$ wavenumbers must lie on the descending branch of the dispersion relation, i.e., $|k|>k_*$ with $k_*=\sqrt{2/\beta}$ for \eqref{eq:disp-trunc} and $k^*=\sqrt{3/(2\beta)}$ for \eqref{eq:disp}. However, it is in this region that both \eqref{eq:disp-trunc} and \eqref{eq:disp} deviate significantly from the shallow-water dispersion relation $\omega_k=|k|$, so these interactions are not physically relevant in the $\beta k^2\ll1$ regime of interest. We refer to these \textit{spurious} solutions and safely discard them in the theoretical and numerical analysis presented in this work. In all cases, we have used  $k_{\mathrm{max}}<k_*$, thus omitting the region in which this interaction is supported.

In summary, the only non-trivial, physically relevant solution for the resonant condition with a concave-down dispersion relation occurs when one wavenumber has an opposite sign to the other three.

\section{Calculation regarding $T_{\mathrm{I}}$ and $T_{\mathrm{II}}$}
\label{appC}
The interaction coefficient \eqref{Tkernek} consists of twelve terms, of which three belong to $T_{\mathrm{I}}$ and nine to $T_{\mathrm{II}}$, according to \eqref{T12}. We begin by evaluating $T_{\mathrm{I}}$, illustrating the procedure with the first term in \eqref{T12}. Substituting \eqref{Vkernel} into \eqref{T12} yields
\begin{equation}
\begin{split}
        -\frac{V_{k,3,k-3}V_{2,1,2-1}}{\omega_1+\omega_{2-1}-\omega_2}=&-\frac{\alpha^2(\omega_k\omega_3\omega_{k-3})^{1/2}(\omega_2\omega_1\omega_{2-1})^{1/2}}{32(\omega_1+\omega_{2-1}-\omega_2)}\\
        &\times[W_{k,3}+W_{k,k-3}+W_{3,k-3}]\\
        &\times[W_{2,1}+W_{2,2-1}+W_{1,2-1}],
\end{split}
\label{ns-1}
\end{equation}
where $W_{i,j}=W(i,j)$. We consider the condition corresponding to configuration 1, where $k,k_1,k_2$ are positive and $k_3$ is negative, with $\kappa_2 > \kappa, \kappa_1, \kappa_3$. The denominator of \eqref{ns-1} can be expanded using \eqref{eq:disp-trunc}, yielding
\[
\begin{split}
\omega_1+\omega_{2-1}-\omega_2=\frac{1}{2}\beta \kappa\kappa_1\kappa_2(\kappa_2-\kappa_1).
\end{split}
\]
where the linear contribution from each term cancels and only $O(\beta)$ terms remain. Substituting this expression into \eqref{ns-1}, and keeping only terms at leading order gives
\begin{equation}
        -\frac{V_{k,3,k-3}V_{2,1,2-1}}{\omega_1+\omega_{2-1}-\omega_2}=\frac{3\alpha^2}{16\beta}\Big(\frac{\kappa_3\kappa}{\kappa_1\kappa_2}\Big)^{1/2}
\label{ns-1-fin}.
\end{equation}
Implementing the same procedure for the remaining $T_{\mathrm{I}}$ terms, we obtain
\begin{equation}
\begin{split}
    -\frac{V_{1,3,1-3}V_{2,k,2-k}}{\omega_k+\omega_{2-k}-\omega_2} =\frac{3\alpha^2}{16\beta}\Big(\frac{\kappa_1\kappa_3}{\kappa\kappa_2}\Big)^{1/2},
\end{split}
\label{ns-2-fin}
\end{equation}
and
\begin{equation}
\begin{split}
-\frac{V_{k+1,k,1}V_{2+3,2,3}}{\omega_{k+1}-\omega_k-\omega_1}=-\frac{3\alpha^2}{16\beta}\Big(\frac{\kappa_3\kappa_2}{\kappa\kappa_1}\Big)^{1/2}.
\end{split}
\label{ns-3-fin}
\end{equation}
By grouping \eqref{ns-1-fin}-\eqref{ns-3-fin} and using $\kappa_3=\kappa_2-\kappa-\kappa_1$, we can obtain a compact expression for $T_{\mathrm{I}}$

\begin{equation}
\begin{split}
    T_{\mathrm{I}}=-\frac{3\alpha^2}{16\beta}\Big(\frac{\kappa_3^3}{\kappa\kappa_1\kappa_2}\Big)^{1/2}.
\end{split}
\label{TI-3}
\end{equation}

We next evaluate the contribution from the $T_{\mathrm{II}}$ terms in \eqref{T12}. The terms in $T_{\mathrm{II}}$ can be further subdivided into $T^a_{\mathrm{II}}$ and $T^{b}_{\mathrm{II}}$ based on how the linear terms in the denominator cancel--those with two linear terms after cancellation grouped into $T^a_{\mathrm{II}}$, and those with one linear term grouped into $T^b_{\mathrm{II}}$. Under this classification, $T^a_{\mathrm{II}}$ includes the first, third, and eleventh term in \eqref{Tkernek}, and $T^b_{\mathrm{II}}$ includes the remaining terms. We begin with an example of the calculation of one term in $T^a_{\mathrm{II}}$, specifically taking the first term in \eqref{Tkernek},
\begin{equation}
\begin{split}
-\frac{V_{k,2,k-2}V_{3,1,3-1}}{\omega_{2}+\omega_{k-2}-\omega_{k}}=&-\frac{\alpha^2(\omega_{k}\omega_{2}\omega_{k-2})^{1/2}(\omega_{3}\omega_1\omega_{3-1})^{1/2}}{32(\omega_{2}+\omega_{k-2}-\omega_{k})}\\
        &\times[W_{k,2}+W_{k,k-2}+W_{2,k-2}][W_{3,1}+W_{3,3-1}+W_{1,3-1}]
\end{split}
\label{reg-1}
\end{equation}
Expanding the denominator and using $\omega_{\kappa-2}=(\kappa_2-\kappa)-\beta/6(\kappa_2-\kappa)^3$ (due to configuration 1) gives
\[
\begin{split}
\omega_{2}+\omega_{k-2}-\omega_k=-2(\kappa_2-\kappa)-\frac{\beta}{6}[\kappa_2^3+(\kappa_2-\kappa)^3-\kappa^3].
\end{split}
\]
Substituting the expression above and only keeping terms at leading order into \eqref{reg-1}, yields
\begin{equation}
\begin{split}
-\frac{V_{k,2,k-2}V_{3,1,3-1}}{\omega_{2}+\omega_{k-2}-\omega_{k}}=-\frac{\alpha^2}{64}(\kappa\kappa_1\kappa_2\kappa_3)^{1/2},
\end{split}
\label{reg-1-fin}
\end{equation}
where the result is only one term after some cancellations. By repeating the same procedure for the remaining two $T^a_{\mathrm{II}}$ terms, one obtains
\begin{equation}
\begin{split}
-\frac{V_{1,2,1-2}V_{3,k,3-k}}{\omega_{2}+\omega_{1-2}-\omega_1}=-\frac{\alpha^2}{64}(\kappa\kappa_1\kappa_2\kappa_3)^{1/2}
\end{split}
\label{reg-2-fin}
\end{equation}
and
\begin{equation}
\begin{split}
-\frac{V_{-k-1,k,1}V_{-2-3,2,3}}{\omega_{k+1}+\omega_{k}+\omega_1} =-\frac{\alpha^2}{64}(\kappa\kappa_1\kappa_2\kappa_3)^{1/2},
\end{split}
\label{reg-3-fin}
\end{equation}.
We next move onto terms belonging to $T^b_{\mathrm{II}}$, using the second term in \eqref{Tkernek} as an example,
\begin{equation}
\begin{split}
-\frac{V_{k,2,k-2}V_{3,1,3-1}}{{\omega_{1}+\omega_{3-1}-\omega_3}}=&-\frac{\alpha^2(\omega_{k}\omega_{2}\omega_{k-2})^{1/2}(\omega_{3}\omega_1\omega_{3-1})^{1/2}}{32({\omega_{1}+\omega_{3-1}-\omega_3})}\\
        &\times[W_{k,2}+W_{k,k-2}+W_{2,k-2}][W_{3,1}+W_{3,3-1}+W_{1,3-1}]
\end{split}
\label{reg-4}
\end{equation}
Expanding the denominator and using $\omega_{3-1}=(\kappa_3+\kappa_1)-\beta/6(\kappa_3+\kappa_1)^2$ (recall configuration 1) reveals that only one linear term remains
\[
\begin{split}
\omega_{1}+\omega_{3-1}-\omega_3=2\kappa_1-\frac{\beta}{6}[\kappa_1^3+(\kappa_3+\kappa_1)^3-\kappa_3^3].
\end{split}
\]
Substituting the above expression into \eqref{reg-4} and retaining only the leading-order contribution gives
\begin{equation}
\begin{split}
-\frac{V_{k,2,k-2}V_{3,1,3-1}}{\omega_{1}+\omega_{3-1}-\omega_3}=-\frac{\alpha^2\kappa_3(\kappa\kappa_1\kappa_2\kappa_3)^{1/2}}{64\kappa_1}-\frac{\alpha^2}{64}(\kappa\kappa_1\kappa_2\kappa_3)^{1/2}.
\end{split}
\label{reg-4-fin}
\end{equation}
Carrying out the same calculation for the five remaining $T^b_{\mathrm{II}}$ gives

\begin{equation}
\begin{split}
-\frac{V_{1,2,1-2}V_{3,k,3-k}}{\omega_{k}+\omega_{3-k}-\omega_3}=-\frac{\alpha^2\kappa_3(\kappa\kappa_1\kappa_2\kappa_3)^{1/2}}{64\kappa}-\frac{\alpha^2}{64}(\kappa\kappa_1\kappa_2\kappa_3)^{1/2},
\end{split}
\label{reg-5-fin}
\end{equation}
\begin{equation}
\begin{split}
-\frac{V_{k,3,k-3}V_{2,1,2-1}}{\omega_3+\omega_{k-3}-\omega_k} =\frac{3\alpha^2\kappa(\kappa\kappa_1\kappa_2\kappa_3)^{1/2}}{64\kappa_3}+\frac{3\alpha^2}{64}(\kappa\kappa_1\kappa_2\kappa_3)^{1/2},
\end{split}
\label{reg-6-fin}
\end{equation}
\begin{equation}
\begin{split}
-\frac{V_{1,3,1-3}V_{2,k,2-k}}{\omega_3+\omega_{1-3}-\omega_1} =\frac{3\alpha^2\kappa_1(\kappa\kappa_1\kappa_2\kappa_3)^{1/2}}{64\kappa_3}+\frac{3\alpha^2}{64}(\kappa\kappa_1\kappa_2\kappa_3)^{1/2},
\end{split}
\label{reg-7-fin}
\end{equation}
\begin{equation}
\begin{split}
-\frac{V_{k+1,k,1}V_{2+3,2,3}}{\omega_{2+3}-\omega_{2}-\omega_3} =\frac{3\alpha^2}{64}(\kappa\kappa_1\kappa_2\kappa_3)^{1/2}-\frac{3\alpha^2\kappa_2(\kappa\kappa_1\kappa_2\kappa_3)^{1/2}}{64\kappa_3},
\end{split}
\label{reg-8-fin}
\end{equation}
and
\begin{equation}
\begin{split}
-\frac{V_{-k-1,k,1}V_{-2-3,2,3}}{\omega_{2+3}+\omega_{2}+\omega_3}=\frac{\alpha^2\kappa_3(\kappa\kappa_1\kappa_2\kappa_3)^{1/2}}{64\kappa_2}
-\frac{\alpha^2}{64}(\kappa\kappa_1\kappa_2\kappa_3)^{1/2}.
\end{split}
\label{reg-9-fin}
\end{equation}
Grouping terms \eqref{reg-1-fin}-\eqref{reg-3-fin} for $T^a_{\mathrm{II}}$ and \eqref{reg-4-fin}-\eqref{reg-9-fin} for $T^b_{\mathrm{II}}$  and using the relation $\kappa_3=\kappa_2-\kappa-\kappa_1$ yields
\begin{equation}
\begin{split}
    T^a_{\mathrm{II}}&=-\frac{3\alpha^2}{64}(\kappa\kappa_1\kappa_2\kappa_3)^{1/2}\\
    T^b_{\mathrm{II}}&=\frac{\alpha^2}{64}(\kappa\kappa_1\kappa_2\kappa_3)^{1/2}\Big[3-\kappa_3\big(\frac{1}{\kappa_1}-\frac{1}{\kappa_2}+\frac{1}{\kappa}\big)\Big].
\end{split}
\end{equation}
Combining $T^a_{\mathrm{II}}$ and $T^b_{\mathrm{II}}$ yields the compact expression

\begin{equation}
    T_{\mathrm{II}}=-\frac{\alpha^2}{64}(\kappa\kappa_1\kappa_2\kappa_3)^{1/2}\Big[\kappa_3\big(\frac{1}{\kappa_1}-\frac{1}{\kappa_2}+\frac{1}{\kappa}\big)\Big].
\label{T2-1}
\end{equation}
Equations \eqref{TI-3} and \eqref{T2-1} are given by \eqref{T12_res} in \S\ref{wke-nf} of the main paper. 

\section{Details of KZ spectra derivation}\label{appD}
\subsection{Reduction of resonance condition in $I$-terms}\label{I-reduct}
We begin by using $I_1$ as an example. To impose the resonant constraint, we substitute the wavenumber resonance condition $\kappa_3=-\kappa-\kappa_1+\kappa_2$ into the frequency resonance condition using \eqref{eq:disp-trunc}, which gives

\[
\omega_\kappa+\omega_1-\omega_2-\omega_3
=-2\kappa_3-\frac{\beta}{6}(\kappa^3+\kappa_1^3-\kappa_2^3-\kappa_3^3)=0.
\]
For a non-trivial solution, $\kappa_3$ must be of $O(\beta)$. As a result, $k_3^3=O(\beta^3)$ is negligible at leading order and can be omitted, yielding
\begin{equation}
\kappa_3=\frac{\beta}{12}\,(\kappa_2^3-\kappa^3-\kappa_1^3).
\label{k3-init}
\end{equation}
The argument of the wavenumber delta function may be written as 
\[
\kappa+\kappa_1-\kappa_2+O(\beta)=0,
\]
since it is appropriate to neglect the $O(\beta)$ term at leading order. This new resonant condition implies $\kappa_2=\kappa+\kappa_1$, which simplifies \eqref{k3-init} to
\begin{equation}
    \kappa_3=\frac{\beta}{4}\kappa\kappa_1\kappa_2.
\label{kappa3}
\end{equation}

Following a similar procedure, we obtain the relations $\kappa_2=(\beta/4)\kappa\kappa_1\kappa_3$ and $\kappa_1=(\beta/4)\kappa\kappa_2\kappa_3$ corresponding to $I_2$ and $I_3$ respectively. Substituting these $\kappa_i\sim\beta$ terms into the interaction coefficient \eqref{Tk123_fin} and integrating out the frequency condition in each case yields \eqref{reduced-I1}-\eqref{reduced-I3} presented in the main paper.
\subsection{Case A}\label{caseA-app}
We seek stationary power-law solutions of the form $n_\kappa\sim\kappa^{-x}$. The total interaction is given by \eqref{total-I} in the main paper, with the kernel $R^\kappa_{12}$ defined by \eqref{int-A}. We adopt the reference form $R^\kappa_{12}\propto(\kappa\kappa_1\kappa_2)^{2-x}\delta(\kappa-\kappa_2-\kappa_1)$ and rewrite the other two permutations, $R^2_{1\kappa}$ and $R^1_{\kappa2}$, into this form via Zakharov transformations (ZT) as introduced in \citet{Zakharov2012}. Inserting the power-law solution ansatz and performing the ZT on $R^2_{1\kappa}$ using $\kappa_1 = \kappa^2/\tilde{\kappa}_1$ and $\kappa_2 = \kappa \tilde{\kappa}_2/\tilde{\kappa}_1$ with Jacobian $|\kappa^3/\tilde{\kappa}_1^3|$ yields
\[
    R^2_{1\kappa}d\kappa_1d\kappa_2\propto(\kappa^{10-4x}\tilde{\kappa}_1^{2x-6}\tilde{\kappa}_2^{2-x})\delta(\kappa-\tilde{\kappa}_2-\tilde{\kappa_1})d\tilde{\kappa_1}d\tilde{\kappa_2}.
\]
From this, the integral contribution relative to the reference kernel becomes
\[
I_2=\int\Big(\frac{\kappa_1}{\kappa}\Big)^{3x-8}R^\kappa_{12}d\kappa_1d\kappa_2,
\]
which serves as a key component for determining the stationary scaling exponent.

Following the same procedure for $R^1_{\kappa2}$ with $\kappa_1=\kappa\tilde{\kappa}_1/\tilde{\kappa_2}$, $\kappa_2=\kappa^2/\tilde{\kappa}_2$ and Jacobian $|\kappa^3/\tilde{\kappa}_2^3|$ yields
\[
    R^1_{\kappa2}d\kappa_1d\kappa_2\propto (\kappa^{10-4x}\tilde{\kappa}_1^{2-x}\tilde{\kappa}_2^{2x-6})\delta(\kappa-\tilde{\kappa}_2-\tilde{\kappa_1})d\tilde{\kappa_1}d\tilde{\kappa_2},
\]
which in turn yield
\[
I_3=\int\Big(\frac{\kappa_2}{\kappa}\Big)^{3x-8}R^\kappa_{12}d\kappa_1\kappa_2.
\]
The total contribution can then be written as
\[I =\int R^\kappa_{12}\Big[1-\Big(\frac{\kappa_1}{\kappa}\Big)^{3x-8}-\Big(\frac{\kappa_2}{\kappa}\Big)^{3x-8}\Big]d\kappa_1d\kappa_2.
\]
which is presented as \eqref{total-I-A} in the main paper.

\subsection{Case B}\label{caseB-app}
Here we consider the solution ansatz $n_\kappa\sim \kappa^{-x}$ and take $R^\kappa_{12}\propto \kappa^{2-2x}\kappa_1^{2-2x}\kappa_2^{2-2x}$ (from \eqref{int-B} in the main paper) as the reference form. Applying the same ZT outlined in \S\ref{caseA-app} to $R^2_{1\kappa}$ yields 
\[
\begin{split}
    R^2_{1\kappa}d\kappa_1d\kappa_2= 4^x\beta^{-x}(\kappa^{10-8x}\tilde{\kappa}_1^{4x-6}\tilde{\kappa}_2^{2-2x})\delta(\kappa-\tilde{\kappa}_2-\tilde{\kappa}_1)d\tilde{\kappa_1}d\tilde{\kappa_2}.
\end{split}
\]
Consequently, the integral contribution relative to the reference kernel is

\[
 I_2=\int\Big(\frac{\kappa_1}{\kappa}\Big)^{6x-8}R^\kappa_{12}d\kappa_1d\kappa_2.
\]
Applying the same procedure to $R^1_{\kappa2}$ using the transformations outlined in \S\ref{caseA-app} gives the following
\[
\begin{split}
    R^1_{\kappa2}d\kappa_1d\kappa_2=4^x\beta^{-x}(\kappa^{10-8x}\tilde{\kappa}_1^{2-2x}\tilde{\kappa}_2^{4x-6})\delta(\kappa-\tilde{\kappa}_2-\tilde{\kappa}_1)d\tilde{\kappa_1}d\tilde{\kappa_2},
\end{split}
\]
which yields
\[
I_3=\Big(\frac{\kappa_2}{\kappa}\Big)
^{6x-8}R^\kappa_{12}d\kappa_1d\kappa_2.
\]
The total interaction term can therefore be written as
\[I =\int R^\kappa_{12}\Big[1-\Big(\frac{\kappa_1}{\kappa}\Big)^{6x-8}-\Big(\frac{\kappa_2}{\kappa}\Big)^{6x-8}\Big]d\kappa_1d\kappa_2,
\]
which corresponds to \eqref{total-I-B} in the main paper.

\subsection{Locality of solutions}\label{app-local}
We analyse the locality of the spectrum $n_\kappa \sim \kappa^{-x}$ by evaluating the behaviour in the infrared (IR) and ultraviolet (UV) limits. 

We begin by assessing the behaviour in the IR limit for Case A. To this end, we take $\kappa_1\rightarrow0$. Expressing the interaction term $R^\kappa_{12}$ given by \eqref{int-A} in terms of the power-law ansatz yields
\begin{equation}
R^\kappa_{12}\propto(\kappa\kappa_1\kappa_2)^2(\kappa^{-x}\kappa_1^{-x}\kappa_2^{-x})\delta(\kappa-\kappa_2-\kappa_1).
\label{Rk12-1}
\end{equation}
Integrating out $\kappa_2$ we obtain
\begin{equation}
\begin{split}
R^{a}(\kappa,\kappa_1)\propto\kappa^{2-x}\kappa_1^{2-x}(\kappa-\kappa_1)^{2-x}
\end{split}
\label{Rk12-2}
\end{equation}
Expanding the $(\kappa-\kappa_1)^{2-x}$ term in \eqref{Rk12-2} leads to
\[
(\kappa-\kappa_1)^{2-x}\approx\kappa^{2-x}-(2-x)\kappa^{1-x}\kappa_1+\dots,
\]
which can then be substituted back into \eqref{Rk12-2}, yielding
\begin{equation}
\begin{split}
R^{a}(\kappa,\kappa_1)\propto\kappa_1^{2-x}\kappa^{4-2x}-(2-x)\kappa^{3-2x}\kappa_1^{3-x}.
\end{split}
\label{Rk12-3}
\end{equation}
Following the same procedure for $R^2_{1\kappa}$, we obtain
\begin{equation}
\begin{split}
R^{b}(\kappa,\kappa_1)\propto\kappa_1^{2-x}\kappa^{4-2x}+(2-x)\kappa^{3-2x}\kappa_1^{3-x}.
\end{split}
\label{R21k-3}
\end{equation}
In the case of $R^1_{\kappa_2}$, taking $\kappa_1\rightarrow0$ while $\kappa$ as $O(1)$ means that the resonance condition cannot be satisfied, therefore $R^c(\kappa,\kappa_1)=0$. From the total interaction integral 
\begin{equation}
\begin{split}
    I(\kappa)&=\int(R^{a}-R^b-R^c)d\kappa_1\sim\int\kappa_1^{3-x}d\kappa_1,
\label{int-R}
\end{split}
\end{equation}
where we see that \eqref{Rk12-3} and \eqref{R21k-3} cancel at the leading order. For the integral \eqref{int-R} to converge, we require $3-x>-1$, resulting in the condition
\begin{equation}
\text{IR convergence (A): }\quad x<4.
\label{IR-conv-A}
\end{equation}

Next, we assess the behaviour in the UV limit for Case A by taking $\kappa_1\rightarrow\infty$. For $R^\kappa_{12}$, the resonance condition cannot be satisfied when taking $\kappa_1\rightarrow\infty$, thus it does not contribute (i.e., $R^a(\kappa,\kappa_1)=0$). For the next term with the solution ansatz
\begin{equation}
    R^2_{1\kappa}\propto(\kappa\kappa_1\kappa_2)^2(\kappa^{-x}\kappa_1^{-x}\kappa_2^{-x})\delta(\kappa_2-\kappa-\kappa_1).
\label{R21k-UV}
\end{equation}
Integrating out $\kappa_2$ yields
\begin{equation}
    R^b(\kappa,\kappa_1)\propto\kappa^{2-x}\kappa_1^{2-x}(\kappa_1+\kappa)^{2-x}
\label{R21k-uv2}
\end{equation}
which, upon substitution of the expansion $(\kappa_1+\kappa)^{2-x}\approx\kappa_1^{2-x}+(2-x)\kappa_1^{1-x}\kappa_1+\dots$ yields
\begin{equation}
    R^b(\kappa,\kappa_1)\propto\kappa^{2-x}\kappa_1^{4-2x}-(2-x)\kappa^{3-x}\kappa_1^{3-2x}.
\label{R21k-uv2}
\end{equation}
Following the the same procedure for $R^1_{\kappa2}$ gives
\begin{equation}
    R^c(\kappa,\kappa_1)\propto\kappa^{2-x}\kappa_1^{4-2x}+(2-x)\kappa^{3-x}\kappa_1^{3-2x}.
\label{R1k2-uv2}
\end{equation}
Explicitly evaluating $R^a-R^b-R^c$ reveals the following scaling in the UV limit
\begin{equation}
\begin{split}
    I(\kappa)\sim\int\kappa_1^{4-2x}d\kappa_1.
\label{int-UV-A}
\end{split}
\end{equation}
Convergence of \eqref{int-UV-A} therefore requires $4-2x<-1$, yielding the criterion
\begin{equation}
\text{UV convergence (A): }\quad x>5/2.
\label{UV-conv-A}
\end{equation}
We next assess the behaviour in the IR limit for Case B by making $\kappa_1\rightarrow0$.  If we express \eqref{int-B} in terms of the power-law ansatz $n_\kappa~\sim\kappa^{-x}$, we obtain
\begin{equation}
\begin{split}
R^\kappa_{12}\propto(\kappa\kappa_1\kappa_2)^2(\kappa^{-x}\kappa_1^{-x}\kappa_2^{-x}\kappa_3^{-x})(\kappa^x-\kappa_1^x-\kappa_2^x)\delta(\kappa-\kappa_2-\kappa_1).
\end{split}
\label{Rk12-B}
\end{equation}
Using $\kappa_3=(\beta/4)\kappa_1\kappa_2\kappa_2$ from \eqref{kappa3}, we can simplify \eqref{Rk12-B} to
\begin{equation}
R^\kappa_{12}\propto(\kappa\kappa_1\kappa_2)^{2-2x}(\kappa^x-\kappa_1^x-\kappa_2^x)\delta(\kappa-\kappa_1-\kappa_2).
\label{Rk12-B2}
\end{equation}
Integrating out $\kappa_2$ gives
\begin{equation}
    R^a(\kappa,\kappa_1)\propto(\kappa\kappa_1(\kappa-\kappa_1))^{2-2x}(\kappa^x+\kappa_1^x-(\kappa-\kappa_1)^x),
    \label{Rk12-B3}
\end{equation}
which upon substituting the expansion of $(\kappa-\kappa_1)^x$ and $(\kappa-\kappa_1)^{2-2x}$ yields
\begin{equation}
R^a(\kappa,\kappa_1)\propto x\kappa^{3-3x}\kappa_1^{3-2x}-(2-2x)x\kappa^{2-3x}\kappa_1^{4-2x}.
\label{Rk12-B3}
\end{equation}
Following the same procedure for $R^2_{1\kappa}$ gives
\begin{equation}
R^b(\kappa,\kappa_1)\propto x\kappa^{3-3x}\kappa_1^{3-2x}+(2-2x)x\kappa^{2-3x}\kappa_1^{4-2x}.
\label{R21k-B3}
\end{equation}
As stated in Case A, $R^1_{\kappa2}$ does not contribute when $\kappa_1\rightarrow0$, therefore $R^c(\kappa,\kappa_1)=0$. Explicitly evaluating $R^a-R^b-R^c$ shows that the leading-order terms of \eqref{Rk12-B3} and \eqref{R21k-B3} cancel. Thus, the integral in the IR limit behaves like
\begin{equation}
I(\kappa)\sim\int\kappa_1^{4-2x}d\kappa_1
\label{int-IR-B}
\end{equation}
For \eqref{int-IR-B} to converge, the exponent must satisfy $4-2x>-1$, which yields the condition
\begin{equation}
\text{IR convergence (B): }\quad x<5/2.
\label{IR-conv-B}
\end{equation}
Next, we take $\kappa_1\rightarrow\infty$ to assess behaviour in the UV limit. In this scenario, $R^{\kappa}_{12}=0$ (and thus $R^a(\kappa,\kappa_1)=0$) for the same reason stated above for Case A. The interaction term $R^{2}_{1\kappa}$ can be expressed as 
\begin{equation}
    R^{2}_{1\kappa}\propto(\kappa\kappa_1\kappa_2)^{2-2x}(\kappa_2^x-\kappa^x-\kappa_1^x)\delta(\kappa_2-\kappa-\kappa_1).
\end{equation}
After integrating out $\kappa_2$ we get
\begin{equation}
    R^b(\kappa,\kappa_1)\propto(\kappa\kappa_1(\kappa_1+\kappa))^{2-2x}(\kappa^x+\kappa_1^x-(\kappa_1+\kappa)^x),
    \label{R21k-B-UV}
\end{equation}
into which we substitute the expansion for $(\kappa_1+\kappa)^{2-2x}$ and $(\kappa_1+\kappa)^{x}$ yielding
\begin{equation}
    R^b(\kappa,\kappa_1)\propto(2x-2x^2)\kappa^{4-2x}\kappa_1^{3-2x}+x\kappa^{3-2x}\kappa_1^{3-3x}.
    \label{R21k-B-simp}
\end{equation}
Carrying out the same procedure on $R^1_{2\kappa}$ gives
\begin{equation}
    R^c(\kappa,\kappa_1)\propto-(2x-2x^2)\kappa^{4-2x}\kappa_1^{3-2x}+x\kappa^{3-2x}\kappa_1^{3-3x}.
    \label{R1k2-B-simp}
\end{equation}
Evaluating $R^a-R^b-R^c$ shows that the first terms in \eqref{R21k-B-simp} and \eqref{R1k2-B-simp} cancel, and the integral scales as
\begin{equation}
I(\kappa)\sim\int\kappa_1^{3-3x}d\kappa_1
\label{int-UV-B}
\end{equation}
 In order for the integral to converge, we require $3-3x<-1$, leading to the condition
\begin{equation}
\text{UV convergence (B): }\quad x>4/3.
\label{UV-conv-B}
\end{equation}
\section{Unidirectional KdV simulations}
\label{appF}
Following the procedure outlined in \citet{Karczewska2018} (see Ch. 3), a KdV equation can be derived from the KB system \eqref{eq:kb1} and \eqref{eq:kb2} by eliminating one direction of wave propagation. The KdV equation reads
\begin{equation}
    \eta_t+\eta_x+\frac{3}{2}\alpha\eta\eta_x+\frac{1}{6}\beta\eta_{xxx}=0.
    \label{kdv-app}
\end{equation}

\begin{figure}
\centerline{\includegraphics{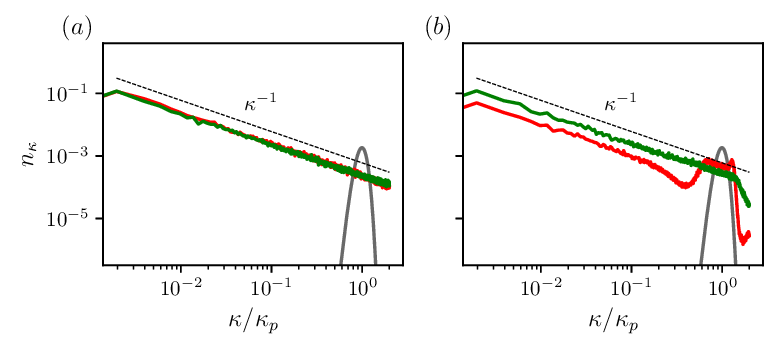}}
\caption{The spectrum $n_k\propto|\eta_k|^2/\omega_k$ at the initial (solid grey) and stationary state for KB (solid green) and KdV (red) for $(a)$ $\alpha=0.2$ taken at $5\times10^5T_p$ and $(b)$ $\alpha=0.02$ taken at $1.2\times10^6T_p$. The thermal-equilibrium scaling is denoted (dashed black).}
\label{fig:kb-kdv}
\end{figure}

We perform numerical simulations of \eqref{kdv-app} using a pseudospectral method combined with an IF-RK4 time-marching scheme. The computational setup follows the configuration detailed in \S\ref{free-evo-config}, with the sole modification that the initial condition now only contains right-propagating waves. Figure \ref{fig:kb-kdv}$(a)$ compares the spectral evolution of the KdV equation and the KB system for $\alpha=\beta=0.2$. Both spectra reach the same thermal-equilibrium form on the same time scale, with the respective spectra evaluated at $t=5\times10^5T_p$. This behaviour is to be expected, as when nonlinearity and dispersion are balanced, the dynamics are primarily governed by quasi-resonant triad interactions, which are supported in both KB and KdV frameworks. 

In contrast, figure \ref{fig:kb-kdv}$(b)$ presents the spectral evolution for $\alpha=0.02$ in the regime $\alpha\ll\beta$, where a clear disparity emerges between the dynamics of the KdV equation and the KB system. We see that the spectrum from the KB equation thermalises at all scales, but the spectrum from the KdV equation only thermalises at large scales. We recall that in this regime of $\alpha\ll\beta$, the normal-form transformation is valid and the quasi-resonant quartet interactions may begin to play a significant role in the spectral evolution. This is indeed the mechanism driving thermalisation of the KB spectrum towards moderate-to-small scales, which, on the other hand, is absent in the KdV system. We further note that the validity condition of the normal-form transformation requires the spectrum to decay sufficiently fast at large scales, which is not the case as in the simulations (more precisely, the validity condition should read $\alpha \ll \beta k^2$, which is not satisfied at small $k$). Therefore, the evolution of large scales may still be driven by triad quasi-resonances in both the KB and KdV simulations, as we observe in figure\ref{fig:kb-kdv}$(b)$.  

\section{Numerical scheme and validation}
\label{appE}
\subsection{Time Integration Scheme}
In this appendix, we describe the IF-RK4 scheme for the KB system. We start by splitting the linear and nonlinear parts of equations \eqref{eq:kb_n1} and \eqref{eq:kb_n2} and expressing them in Fourier space (essentially following the procedure in \citet{Pan2020}):
\begin{equation}
    \frac{\partial}{\partial t} 
    \begin{bmatrix}
    \hat{\eta}_k \\
    \hat{u}_k
\end{bmatrix}
= ik
    \begin{bmatrix}
    0 & -1+\frac{\beta}{3}k^2 \\
    -1 & 0 
\end{bmatrix}
    \begin{bmatrix}
    \hat{\eta}_k \\
    \hat{u}_k
\end{bmatrix} +
    \begin{bmatrix}
    \hat{P}_k \\
    \hat{Q}_k
\end{bmatrix}
\label{eq:mat_eq}
\end{equation}
where $\hat{\eta}_k$ and $\hat{u}_k$ are the Fourier transforms of $\eta(x)$ and $u(x)$, and $\hat{P}_k$ and $\hat{Q}_k$ represent the nonlinear terms in \eqref{eq:kb_n1} and \eqref{eq:kb_n2}.
Diagonalising \eqref{eq:mat_eq} yields
\begin{equation}
    \frac{\partial}{\partial t} 
    \begin{bmatrix}
    v_1 \\
    v_2
\end{bmatrix}
= ik
    \begin{bmatrix}
    \lambda_1 & 0 \\
    0 & \lambda_2 
\end{bmatrix}
    \begin{bmatrix}
    v_1 \\
    v_2
\end{bmatrix} +
    \begin{bmatrix}
    F_1 \\
    F_2
\end{bmatrix}
\label{eq:diag_eq}
\end{equation}
where $\lambda_1=-ik\sqrt{1-\frac{\beta}{3}k^2}$ and $\lambda_2=ik\sqrt{1-\frac{\beta}{3}k^2}$, and
\begin{equation}
    \begin{bmatrix}
    v_1 \\
    v_2
\end{bmatrix}
= \Omega^{-1}
    \begin{bmatrix}
    \hat{\eta}_k \\
    \hat{u}_k
\end{bmatrix},
\label{eq:v1v2}
\end{equation}
\begin{equation}
    \begin{bmatrix}
    F_1 \\
    F_2
\end{bmatrix}
= \Omega^{-1}
    \begin{bmatrix}
    \hat{P}_k \\
    \hat{Q}_k
\end{bmatrix},
\label{eq:F1F2}
\end{equation}
with
\begin{equation}
    \Omega = \begin{bmatrix}
    \sqrt{1-\frac{\beta}{3}k^2} & - \sqrt{1-\frac{\beta}{3}k^2} \\
    1 & 1
\end{bmatrix}.
\end{equation}
We then define $\Psi_1=\exp(-\lambda_1t)v_1$ and $\Psi_2=\exp(-\lambda_2t)v_2$, with which \eqref{eq:diag_eq} can be transformed into
\begin{equation}
    \frac{\partial}{\partial t}\Psi_1=\exp(-\lambda_1t)F_1,
\label{dtpsi1}
\end{equation}
\begin{equation}
    \frac{\partial}{\partial t}\Psi_2=\exp(-\lambda_2t)F_2.
\label{dtpsi2}
\end{equation}
Equations \eqref{dtpsi1} and \eqref{dtpsi2} can then be integrated using the standard RK4 scheme. This combined IF-RK4 scheme essentially solves the linear part of \eqref{eq:diag_eq} analytically (to machine precision, see figure \ref{fig:kb_lin} for validation). For the non-integrable Boussinesq variant, we modify the linear operator in \eqref{eq:mat_eq} to yield $\omega_k=k(1-(\beta/6)k^2)$, with eigenvalues $\lambda_1=-i\omega_k$ and $\lambda_2=i\omega_k$. The diagonalisation and IF-RK4 update are then applied exactly as above. A similar scheme is implemented for simulations of the KdV equation. An integrating factor can be applied directly to the KdV equation \citep{Treferton2000}, in contrast to the coupled KB system which must first be diagonalised. 

For both KB and KdV, the spatial derivatives, including the nonlinear terms $\hat{P}_k$ and $\hat{Q}_k$ in the former, are computed using Fast Fourier Transforms.  To mitigate aliasing errors, we use a 1/2 de-aliasing rule.

\subsection{Validation of KB System}\label{kb-valid}
\begin{figure}
\centerline{\includegraphics[]{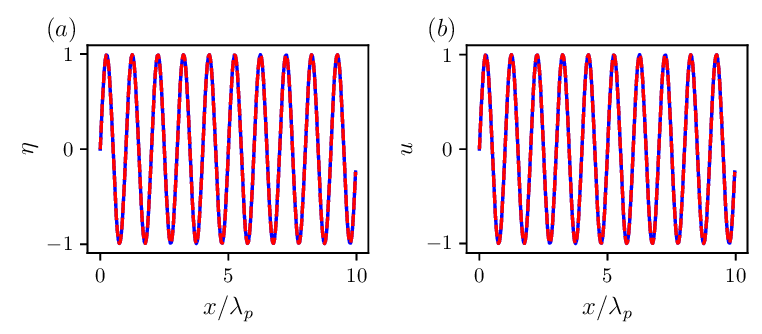}}
\caption{Comparison of the numerical (solid blue) and analytical (dashed red) solutions for $(a)$ surface elevation and $(b)$ velocity for the linear portion of the KB system at $t=10T_p$.}
\label{fig:kb_lin}
\end{figure}

We perform a series of tests to validate the aforementioned methods. We begin by ensuring that the linear part of the KB system is solved to machine precision using the integrating factor scheme. To do this, we set $\alpha=0$, discarding the nonlinear terms in \eqref{eq:kb_n1} and \eqref{eq:kb_n2}, we are left with:

\begin{equation}
    \eta_t =- u_x -\frac{1}{3}\beta u_{xxx},
\label{eq:kb_l1}
\end{equation}
\begin{equation}
    u_t=-\eta_x.
\label{eq:kb_l2}
\end{equation}
This system has an exact solution of the form $\sin(kx-\omega t)$. Substituting this in for $u$, we get:
\begin{equation}
\begin{split}
    u(x,t)=\sin(kx-\omega t)\\
    \eta(x,t)=\frac{\omega}{k}\sin(kx-\omega t)
\label{lin-kb}
\end{split}
\end{equation}
where the dispersion relation is given by \eqref{eq:disp}. For this test case, we set $\beta=1$ and use \eqref{lin-kb} as our initial condition with $t=0$. Figure \ref{fig:kb_lin} shows the numerical solution compared against the analytic solution after $10T_p$. Both the velocity component and the surface elevation are within machine precision of the analytic solution. 

\begin{figure}
\centerline{\includegraphics[]{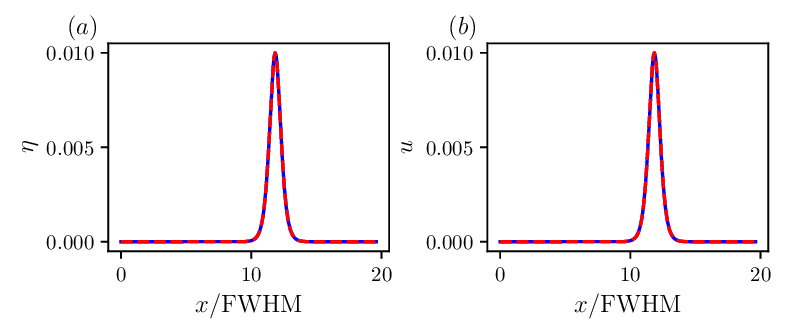}}
\caption{Comparison of the numerical (solid blue) and analytical (dashed red) solutions for $(a)$ surface elevation and $(b)$ velocity for the full KB system at $t=2\tau_s$.}
\label{fig:kb_nlin}
\end{figure}

\begin{figure}
\centerline{\includegraphics[]{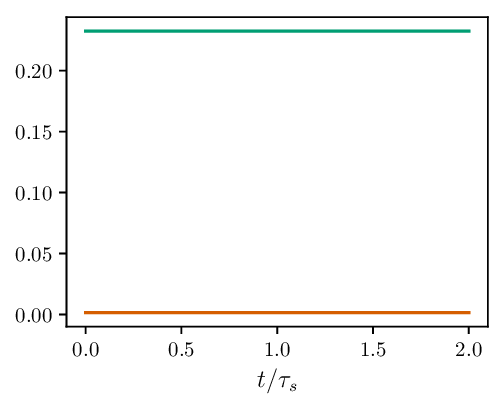}}
\caption{Total momentum $P$ (green line) and energy $H$ (orange line) plotted at each time step over the duration of $2\tau_s$ for the full KB system.}
\label{fig:kb_cons}
\end{figure}

We proceed with the full KB system where we set $\alpha=\beta=1$. The solver is tested using a known exact soliton solution (example taken from \citet{Juliussen2014})
\begin{equation}
\begin{split}
    u(x,t)=\frac{2(c^2-1)}{\cosh(\sqrt{3(c^2-1)}(x-ct))+c} \\
    \eta(x,t)=cu(x,t)-\frac{1}{2}u^2(x,t)
\end{split}    
\end{equation}
with $c=1.005$. We characterise the time scale of the system by the pulse duration $\tau_s=\mathrm{FWHM}/c$ where FWHM is the full width at half-maximum for the soliton profile and $c$ is the soliton velocity. We allow the soliton to propagate for $2\tau_s$. Figure \ref{fig:kb_nlin} shows a comparison between the numerical and analytical solutions at $2\tau_s$, demonstrating the high accuracy of the numerical results. 
We test the conservation of two quantities, the total energy $H$ and the total momentum $P$, defined by the integrals
\begin{equation}
    H(t)=\int E(x,t)dx,\qquad P(t)=\int I(x,t)dx.
\label{cons}
\end{equation}
For the KB system, the energy density $E$ is expressed as
\begin{equation}
E = \frac{\alpha^2}{2} \eta^2 + \frac{\alpha^2}{2} (1 + \alpha \eta) w^2 + \frac{\alpha^2 \beta}{3} w w_{xx} + \frac{\alpha^2 \beta}{6} w_x^2,
\end{equation}
and the momentum density $I$ takes the form 
\begin{equation}
I = \alpha w + \alpha^2 w \eta + \frac{1}{3} \alpha \beta w_{xx},
\end{equation}
as presented in \citet{Ali2017}. Figure \ref{fig:kb_cons} demonstrates that the energy $H$ and the momentum $P$ remain constant for the duration of the simulation.
\subsection{Validation of KdV Equation}
\begin{figure}
\centerline{\includegraphics[]{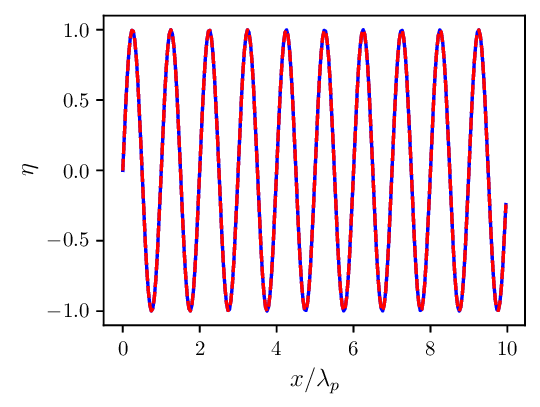}}
\caption{Comparison of the numerical (solid blue) and analytical (red dashed line) solutions for surface elevation for the linear portion of the KdV equation at $t=10T_p$.}
\label{fig:kdv_lin}
\end{figure}

\begin{figure}
\centerline{\includegraphics[]{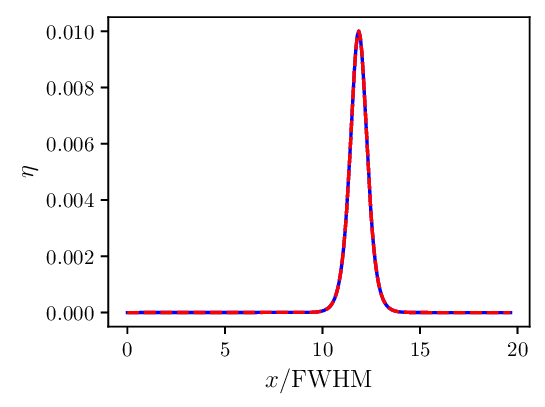}}
\caption{Comparison of the numerical (solid blue) and analytical (dashed red) solutions for surface elevation for the full KdV equation at $t=2\tau_s$.}
\label{fig:kdv_nlin}
\end{figure}

\begin{figure}
\centerline{\includegraphics[]{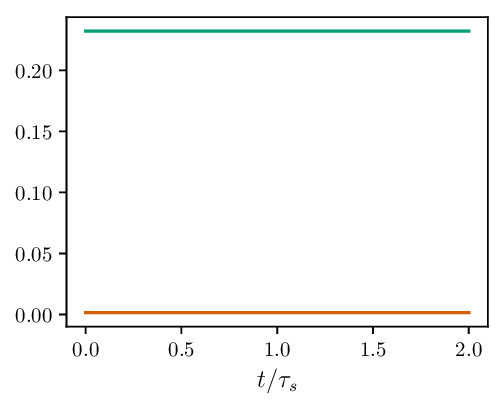}}
\caption{Total momentum $P$ (green line) and energy $H$ (orange line) plotted at each time step over the duration of $2\tau_s$ for the full KdV equation.}
\label{fig:kdv_cons}
\end{figure}

We begin by considering the linear part of the KdV equation. Setting $\alpha=0$ in \eqref{kdv-app} leaves
\begin{equation}
        \eta_t+\eta_x+\frac{1}{6}\beta\eta_{xxx}=0,
    \label{linear-kdv}
\end{equation}
which has an analytical solution of the form $\eta(x,t)=\sin(kx-wt)$ with $\omega=k-\frac{\beta}{6}k^3$. For this test case, we set $\beta=1$ and use \eqref{linear-kdv} with $t=0$ as our initial condition. Figure \ref{fig:kdv_lin} compares the numerical and analytical solutions after $10T_p$. As expected from the integrating factor scheme used, the surface elevation agrees with the analytical result to machine precision.
Proceeding with the full KdV equation, we set $\alpha=\beta=1$ and take a known soliton solution 
\begin{equation}
\eta(x,t)=2(c-1)\sech^2 \Bigg[\sqrt{\frac{3}{2}(c-1)}\bigg(x-ct)\bigg)\Bigg]
\end{equation}
with $c=1.005$ and $t=0$ for our initial condition. We follow the same procedure as in \ref{kb-valid} by defining the time scale in terms of $\tau_s$. Figure \ref{fig:kdv_nlin} illustrates the high degree of precision between the numerical and analytical solutions.
Conservation of total energy $H$ and total momentum $E$ for KdV is also validated using the integrals defined in \eqref{cons}; however, in this case, the energy density is defined as
\begin{equation}
E = \alpha^2 \eta^2 + \frac{\alpha^3}{4} \eta^3 + \frac{\alpha^2 \beta}{6} \eta \eta_{xx} + \frac{\alpha^2 \beta}{6} \eta_x^2,
\end{equation}
while the momentum density is 
\begin{equation}
I = \alpha \eta + \frac{3}{4} \alpha^2 \eta^2 + \frac{1}{6} \alpha \beta \eta_{xx},
\end{equation}
as derived in \citet{Ali2013}. Figure \ref{fig:kdv_cons} confirms the conservation of both energy $H$ and momentum $P$ for the duration of the simulation.

\section{Numerical scheme for DST}
\label{appG}

We solve \eqref{spec-zeta} using a finite-difference method on a uniform grid $x_j=-L/2+j\Delta x$, $j=0,\dots,N-1$ and $\Delta x =L/(N-1)$. Defining the grid values by $\Psi_j = \Psi(x_j)$, we approximate $\Psi_{xx}(x_j)$ at the interior points $j=1,\dots,N-2$ using a second-order central difference stencil

\begin{equation}
\frac{\Psi_{j-1}-2\Psi_j+\Psi_{j+1}}{\Delta x^2}+V_j\Psi_j=0
\label{FD-int}
\end{equation}
where $V_j \coloneq E(\zeta)^2+ik(\zeta)q(x_j)+r(x_j)$. To capture bound states (elements of the discrete spectrum), we impose asymptotically decaying boundary conditions. That is, for sufficiently large $|x|$
\[
\Psi(x)\sim e^{(-iE|x|)},\qquad \mathrm{Im}[E]>0,
\]
so that
\[
\Psi_x(x)\sim -iE\;\mathrm{sgn}(x)\Psi(x).
\]
The requirement of $\mathrm{Im}[E]>0$ places us in the upper half $E$-plane, which ensures decay. These asymptotics lead to Robin boundary conditions
\[
\Psi_x(-L/2)-iE\Psi(-L/2)=0,\ \  \Psi_x(L/2)+iE\Psi(L/2)=0.
\]
These boundary conditions are applied with second-order one-sided finite differences at the endpoints. At the left boundary, we use the forward stencil
\begin{equation}
\frac{-3\Psi_0+4\Psi_1-\Psi_2}{2\Delta x}-iE\Psi_0=0,
\label{FD-left}
\end{equation}
and at the right boundary we use the backward stencil
\begin{equation}
\frac{3\Psi_{N-1}-4\Psi_{N-2}+\Psi_{N-3}}{2\Delta x}+iE\Psi_{N-1}=0.
\label{FD-right}
\end{equation}

By assembling \eqref{FD-int}, \eqref{FD-left}, \eqref{FD-right} we obtain a $N\times N$ linear system $A(\zeta)\Psi=0$. The majority of the matrix $A(\zeta)$ consists of tridiagonal entries corresponding to the interior discretisation, while the boundary conditions are incorporated by modifying the matrix's first and last rows. Bound states correspond to values of $\zeta$ for which $A(\zeta)$ is singular, i.e.,
\[
\mathrm{det}A(\zeta)=0,
\]
defining a nonlinear eigenvalue problem. Instead of evaluating $\mathrm{det}A(\zeta)$ directly, we locate roots via a Newton iteration of the smallest singular value of $A(\zeta)$ (as zeros of the determinant correspond to points where the smallest singular value is zero). At each iterate $\zeta$, the singular value decomposition $A(\zeta)=U\Sigma V^*$ is computed. From this, we extract the minimal singular value $\sigma\coloneqq\sigma_{\mathrm{min}}(A(\zeta))$ and its corresponding left and right singular vectors $u$ and $v$, respectively. We can then define the derivative as $d\sigma/d\zeta=u^*A'(\zeta)v$ where $A'(\zeta)$ is simply the derivative of the matrix $A(\zeta)$ with respect to $\zeta$ \citep{Chen2018}. The update formula is given by
\[
d\zeta=-\frac{\sigma}{u^*A'(\zeta)v}
\]
so that $\zeta\leftarrow\zeta+d\zeta$ at each iteration. We iterate until $\sigma$ falls below some prescribed tolerance, indicating that a root has been located. To ensure robust identification of all roots, we implement a multi-start Newton's method by selecting the initial value of $\zeta$ from a sufficiently fine mesh spanning the complex $\zeta$-plane. We remark that for $\beta>0$, bound states correspond either to purely imaginary eigenvalues ($\zeta=i\eta$, $\eta\in\mathbb{R}$), or to complex-conjugate pairs ($\zeta_2=-\zeta_1^*$), as in the case of breathers in the Sine-Gordon equation \citep{Kaup1975}. 

The criterion for the robustness of bound states corresponding to solitons can be stated in terms of the soliton velocity. The velocity profile for a KB soliton is given by
\begin{equation}
u(x,t)=\frac{2(c_s^2-1)}{\alpha\ \mathrm{cosh}(\sqrt{\frac{3}{\beta}(c_s^2-1)}(x-c_st))+c_s},
\label{u-sol}
\end{equation}
where $c_s$ is the soliton velocity, which can be computed via the discrete eigenvalues $\zeta$ \citep{Kaup1975}. To ensure physical realisability, the soliton velocity must satisfy $|c_s|>1$. This condition ensures that the quantity $K$ corresponding to $\sqrt{\frac{3}{\beta}(c_s^2-1)}$ appearing in the hyperbolic cosine argument of \eqref{u-sol} remains real and positive. The parameter $K$ controls the inverse width of the soliton profile, i.e., width $\propto 1/K$; therefore, solitons with velocities closer to 1 (corresponding to eigenvalues closer to the continuous spectrum) exhibit broad profiles with slow decay in the physical domain. These solitons are not expected to be realised in our finite domain due to their excessive width. Solitons with larger velocities correspond to narrower, more strongly localised solitons. This is consistent with the definition of ``true" bound states where the tails decay to zero as $|x|\rightarrow\infty$. In the $\zeta$-plane, the continuous spectrum corresponds to $\textnormal{Im}[E(\zeta)]=0$, i.e., $\zeta\in\mathbb{R}$ or $|\zeta|=\sqrt{3/B}$. We therefore retain only discrete eigenvalues satisfying $|c_s|>1$ and separated from the continuous spectrum curve by at least a \%1 buffer, $\big||\zeta|-\sqrt{3/\beta}\big|>0.01\sqrt{3/\beta}$. Any discrete eigenvalue that does not satisfy this condition is deemed spurious and is not included in the computation of the soliton energy presented in \S\ref{free-evo}.
\bibliographystyle{jfm}
\bibliography{jfm}

@article {Janssen2003,
      author = "Peter A. E. M. Janssen",
      title = {{Nonlinear Four-Wave Interactions and Freak Waves}},
      journal = "Journal of Physical Oceanography",
      year = "2003",
      publisher = "American Meteorological Society",
      address = "Boston MA, USA",
      volume = "33",
      number = "4",
      doi = "10.1175/1520-0485(2003)33<863:NFIAFW>2.0.CO;2",
      pages=      "863 - 884",
      url = "https://journals.ametsoc.org/view/journals/phoc/33/4/1520-0485_2003_33_863_nfiafw_2.0.co_2.xml"
}

@article{Annenkov2006, title={{Role of non-resonant interactions in the evolution of nonlinear random water wave fields}}, volume={561}, DOI={10.1017/S0022112006000632}, journal={Journal of Fluid Mechanics}, author={Annenkov, Sergei Y. and Shrira, Victor I.}, year={2006}, pages={181–207}}

@article{Bona2004,
doi = {10.1088/0951-7715/17/3/010},
url = {https://doi.org/10.1088/0951-7715/17/3/010},
year = {2004},
month = {feb},
publisher = {},
volume = {17},
number = {3},
pages = {925},
author = {J L Bona and M Chen and J-C Saut},
title = {{Boussinesq equations and other systems for small-amplitude long waves in nonlinear dispersive media: II. The nonlinear theory}},
journal = {Nonlinearity}
}

@book{Zakharov1991, address={Berlin, Heidelberg}, series={Springer Series in Nonlinear Dynamics}, title={{What Is Integrability?}},author = {V.E. Zakharov}, rights={http://www.springer.com/tdm}, ISBN={9783642887055}, url={http://link.springer.com/10.1007/978-3-642-88703-1}, DOI={10.1007/978-3-642-88703-1},   number ={},publisher={Springer Berlin Heidelberg}, year={1991}, collection={Springer Series in Nonlinear Dynamics} }

@article{Zakharov1980,
title = {{Degenerative dispersion laws, motion invariants and kinetic equations}},
journal = {Physica D: Nonlinear Phenomena},
volume = {1},
number = {2},
pages = {192-202},
year = {1980},
issn = {0167-2789},
doi = {https://doi.org/10.1016/0167-2789(80)90011-1},
url = {https://www.sciencedirect.com/science/article/pii/0167278980900111},
author = {V.E. Zakharov and E.I. Schulman}
}

@article{Zakharov1988,
title = {{On additional motion invariants of classical Hamiltonian wave systems}},
journal = {Physica D: Nonlinear Phenomena},
volume = {29},
number = {3},
pages = {283-320},
year = {1988},
issn = {0167-2789},
doi = {https://doi.org/10.1016/0167-2789(88)90033-4},
url = {https://www.sciencedirect.com/science/article/pii/0167278988900334},
author = {V.E. Zakharov and E.I. Schulman}
}

@article{Klein2025, title={{On the Kaup–Broer–Kupershmidt systems}}, volume={12}, ISSN={2308-2151}, url={https://ems.press/journals/emss/articles/14298696}, DOI={10.4171/emss/98}, abstractNote={Christian Klein, Jean-Claude Saut}, number={1}, journal={EMS Surveys in Mathematical Sciences}, author={Klein, Christian and Saut, Jean-Claude}, year={2025}, month=apr, pages={215–242}}

@article{Bona2002, title={{Boussinesq Equations and Other Systems for Small-Amplitude Long Waves in Nonlinear Dispersive Media. I: Derivation and Linear Theory}}, volume={12}, ISSN={1432-1467}, url={https://doi.org/10.1007/s00332-002-0466-4}, DOI={10.1007/s00332-002-0466-4}, abstractNote={Considered herein are a number of variants of the classical Boussinesq system and their higher-order generalizations. Such equations were first derived by Boussinesq to describe the two-way propagation of small-amplitude, long wavelength, gravity waves on the surface of water in a canal. These systems arise also when modeling the propagation of long-crested waves on large lakes or the ocean and in other contexts. Depending on the modeling of dispersion, the resulting system may or may not have a linearization about the rest state which is well posed. Even when well posed, the linearized system may exhibit a lack of conservation of energy that is at odds with its status as an approximation to the Euler equations. In the present script, we derive a four-parameter family of Boussinesq systems from the two-dimensional Euler equations for free-surface flow and formulate criteria to help decide which of these equations one might choose in a given modeling situation. The analysis of the systems according to these criteria is initiated.}, number={4}, journal={Journal of Nonlinear Science}, author={Bona and Chen and Saut}, year={2002}, month=aug, pages={283–318}}

@article{Pan2017, title={{Understanding discrete capillary-wave turbulence using a quasi-resonant kinetic equation}}, volume={816}, ISSN={0022-1120, 1469-7645}, url={https://www.cambridge.org/core/journals/journal-of-fluid-mechanics/article/abs/understanding-discrete-capillarywave-turbulence-using-a-quasiresonant-kinetic-equation/765E14A9DE0202A2A50B13332B2992A2}, DOI={10.1017/jfm.2017.106}, abstractNote={Experimental and numerical studies have shown that, with sufficient nonlinearity, the theoretical capillary-wave power-law spectrum derived from the kinetic equation (KE) of weak turbulence theory can be realized. This is despite the fact that the KE is derived assuming an infinite domain with continuous wavenumber, while experiments and numerical simulations are conducted in realistic finite domains with discrete wavenumbers for which the KE theoretically allows no energy transfer. To understand this, we first analyse results from direct simulations of the primitive Euler equations to elucidate the role of nonlinear resonance broadening (NRB) in discrete turbulence. We define a quantitative measure of the NRB, explaining its dependence on the nonlinearity level and its effect on the properties of the obtained stationary power-law spectra. This inspires us to develop a new quasi-resonant kinetic equation (QKE) for discrete turbulence, which incorporates the mechanism of NRB, governed by a single parameter  $unicode[STIX]{x1D705}$  expressing the ratio of NRB and wavenumber discreteness. At  $unicode[STIX]{x1D705}=unicode[STIX]{x1D705}_{0}approx 0.02$ , the QKE recovers simultaneously the spectral slope  $unicode[STIX]{x1D6FC}_{0}=-17/4$  and the Kolmogorov constant  $C_{0}=6.97$  (corrected from the original derivation) of the theoretical continuous spectrum, which physically represents the upper bound of energy cascade capacity for the discrete turbulence. For  $unicode[STIX]{x1D705}<unicode[STIX]{x1D705}_{0}$ , the obtained spectra represent those corresponding to a finite domain with insufficient nonlinearity, resulting in a steeper spectral slope  $unicode[STIX]{x1D6FC}<unicode[STIX]{x1D6FC}_{0}$  and reduced capacity of energy cascade  $C>C_{0}$ . The physical insights from the QKE are corroborated by direct simulation results of the Euler equations.}, journal={Journal of Fluid Mechanics}, author={Pan, Yulin and Yue, Dick K. P.}, year={2017}, month=apr, pages={R1} }

@article{Zakharov1999,
title = {{Statistical theory of gravity and capillary waves on the surface of a finite-depth fluid}},
journal = {European Journal of Mechanics - B/Fluids},
volume = {18},
number = {3},
pages = {327-344},
year = {1999},
note = {Three-Dimensional Aspects of Air-Sea Interaction},
issn = {0997-7546},
doi = {https://doi.org/10.1016/S0997-7546(99)80031-4},
url = {https://www.sciencedirect.com/science/article/pii/S0997754699800314},
author = {V Zakharov}
}

@book{Zakharov2012,
  title={{Kolmogorov spectra of turbulence I: Wave turbulence}},
  author={Zakharov, Vladimir E and L'vov, Victor S and Falkovich, Gregory},
  year={2012},
  publisher={Springer Science \& Business Media}
}

@article{Onorato2009,
  title={{Four-wave resonant interactions in the classical quadratic Boussinesq equations}},
  author={Onorato, Miguel and Osborne, Alfred Richard and Janssen, PAEM and Resio, D},
  journal={Journal of Fluid Mechanics},
  volume={618},
  pages={263--277},
  year={2009},
  publisher={Cambridge University Press}
}

@article{Kaihatu2007,
  title={{Asymptotic behavior of frequency and wave number spectra of nearshore shoaling and breaking waves}},
  author={Kaihatu, James M and Veeramony, Jayaram and Edwards, Kacey L and Kirby, James T},
  journal={Journal of Geophysical Research: Oceans},
  volume={112},
  number={C6},
  year={2007},
  publisher={Wiley Online Library}
}

@article{Pan2020,
title = {{High-order spectral method for the simulation of capillary waves with complete order consistency}},
journal = {Journal of Computational Physics},
volume = {408},
pages = {109299},
year = {2020},
issn = {0021-9991},
doi = {https://doi.org/10.1016/j.jcp.2020.109299},
url = {https://www.sciencedirect.com/science/article/pii/S0021999120300735},
author = {Yulin Pan}
}

@book{Treferton2000,
author = {Trefethen, Lloyd N.},
title = {{Spectral Methods in MATLAB}},
publisher = {Society for Industrial and Applied Mathematics},
year = {2000},
doi = {10.1137/1.9780898719598},
address = {},
edition   = {}
}

@mastersthesis{Juliussen2014, type={Master thesis}, title={{Investigations of the Kaup-Boussinesq model equations for water waves}}, rights={Copyright the author. All rights reserved}, url={https://bora.uib.no/bora-xmlui/handle/1956/8335}, abstractNote={The Kaup-Boussinesq system is a coupled system of nonlinear partial differential equations which has been derived as a model for surface waves in the context of the Boussinesq scaling, and it has also been derived for an internal wave system. In this thesis, modeling properties of the Kaup-Boussinesq water-wave model are under investigation. Differential balance laws for mass, momentum and energy are considered, and we present an exact differential balance for momentum. A Kaup-Boussinesq system describing long internal waves is investigated and compared with the Gardner equation. Finally, a spectral method for the numerical discretization of the Kaup-Boussinesq system for surface waves is put forward, and shown to converge and be stable.}, school={The University of Bergen}, author={Juliussen, Bjørn-Sverre Hjelle}, year={2014}, month=jun}

@article{Ali2017,
	author = {{Ali, A.} and {Juliussen, B.-S.} and {Kalisch, H.}},
	title = {{Approximate Conservation Laws for an Integrable Boussinesq System}},
	DOI= "10.1051/mmnp/201712101",
	url= "https://doi.org/10.1051/mmnp/201712101",
	journal = {Math. Model. Nat. Phenom.},
	year = 2017,
	volume = 12,
	number = 1,
	pages = "1-14",
}

@article{Ali2013,
  title = {{On the Formulation of Mass,  Momentum and Energy Conservation in the KdV Equation}},
  volume = {133},
  ISSN = {1572-9036},
  url = {http://dx.doi.org/10.1007/s10440-013-9861-0},
  DOI = {10.1007/s10440-013-9861-0},
  number = {1},
  journal = {Acta Applicandae Mathematicae},
  publisher = {Springer Science and Business Media LLC},
  author = {Ali,  Alfatih and Kalisch,  Henrik},
  year = {2013},
  month = dec,
  pages = {113–131}
}

@article{Kaup1975,
    author = {Kaup, D. J.},
    title = {{A Higher-Order Water-Wave Equation and the Method for Solving It}},
    journal = {Progress of Theoretical Physics},
    volume = {54},
    number = {2},
    pages = {396-408},
    year = {1975},
    month = {08},
    issn = {0033-068X},
    doi = {10.1143/PTP.54.396},
    url = {https://doi.org/10.1143/PTP.54.396},
}

@book{Karczewska2018, place={Zielona Góra}, title={{Shallow water waves: Extended Kortweg-de Vries equations: Second order perturbation approach}}, publisher={Oficyna Wydawnicza Uniwersytetu Zielonogórskiego}, author={Karczewska, Anna and Rozmej, Piotr}, year={2018}}

@article{Bhrawy2013, title={Integrable system modelling shallow water waves: Kaup–Boussinesq shallow water system}, volume={87}, rights={http://www.springer.com/tdm}, ISSN={0973-1458, 0974-9845}, url={http://link.springer.com/10.1007/s12648-013-0260-1}, DOI={10.1007/s12648-013-0260-1}, number={7}, journal={Indian Journal of Physics}, author={Bhrawy, A. H. and Tharwat, M. M. and Abdelkawy, M. A.}, year={2013}, month=jul, pages={665–671} }

@article{Gong2022, title={{Formation of the undular bores in shallow water generalized Kaup–Boussinesq model}}, volume={439}, ISSN={01672789}, url={https://linkinghub.elsevier.com/retrieve/pii/S0167278922001506}, DOI={10.1016/j.physd.2022.133398}, journal={Physica D: Nonlinear Phenomena}, author={Gong, Ruizhi and Wang, Deng-Shan}, year={2022}, month=nov, pages={133398}}

@article{Ivanov2009, title={{Two-component integrable systems modelling shallow water waves: The constant vorticity case}}, volume={46}, rights={https://www.elsevier.com/tdm/userlicense/1.0/}, ISSN={01652125}, url={https://linkinghub.elsevier.com/retrieve/pii/S0165212509000493}, DOI={10.1016/j.wavemoti.2009.06.012}, number={6}, journal={Wave Motion}, author={Ivanov, Rossen}, year={2009}, month=sep, pages={389–396} }

@article{Kupershmidt1985, title={{Mathematics of dispersive water waves}}, volume={99}, rights={http://www.springer.com/tdm}, ISSN={0010-3616, 1432-0916}, url={http://link.springer.com/10.1007/BF01466593}, DOI={10.1007/BF01466593}, number={1}, journal={Communications in Mathematical Physics}, author={Kupershmidt, B. A.}, year={1985}, month=mar, pages={51–73}}

@article{El2001, title={{Integrable Shallow‐Water Equations and Undular Bores}}, volume={106}, rights={http://onlinelibrary.wiley.com/termsAndConditions#vor}, ISSN={0022-2526, 1467-9590}, url={https://onlinelibrary.wiley.com/doi/10.1111/1467-9590.00163}, DOI={10.1111/1467-9590.00163}, abstractNote={On the basis of the integrable Kaup–Boussinesq version of the shallow‐water equations, an analytical theory of undular bores is constructed. A complete classification for the problem of the decay of an initial discontinuity is made.}, number={2}, journal={Studies in Applied Mathematics}, author={El, G. A. and Grimshaw, R. H. J. and Pavlov, M. V.}, year={2001}, month=feb, pages={157–186} }

@article{Hasselmann1962, title={{On the non-linear energy transfer in a gravity-wave spectrum Part 1. General theory}}, volume={12}, rights={https://www.cambridge.org/core/terms}, ISSN={0022-1120, 1469-7645}, url={https://www.cambridge.org/core/product/identifier/S0022112062000373/type/journal_article}, DOI={10.1017/S0022112062000373}, abstractNote={The energy flux in a finite-depth gravity-wave spectrum resulting from weak non-linear couplings between the spectral components is evaluated by means of a perturbation method. The fifth-order analysis yields a fourth-order effect comparable in magnitude to the generating and dissipating processes in wind-generated seas. The energy flux favours equidistribution of energy and vanishes in the limiting case of a white, isotropic spectrum. The influence on the equilibrium structure of fully developed wave spectra and on other phenomena in random seas is discussed briefly.}, number={4}, journal={Journal of Fluid Mechanics}, author={Hasselmann, K.}, year={1962}, month=apr, pages={481–500}}

@article{Redor2019,
  title = {{Experimental Evidence of a Hydrodynamic Soliton Gas}},
  author = {Redor, Ivan and Barth\'elemy, Eric and Michallet, Herv\'e and Onorato, Miguel and Mordant, Nicolas},
  journal = {Phys. Rev. Lett.},
  volume = {122},
  issue = {21},
  pages = {214502},
  numpages = {6},
  year = {2019},
  month = {May},
  publisher = {American Physical Society},
  doi = {10.1103/PhysRevLett.122.214502},
  url = {https://link.aps.org/doi/10.1103/PhysRevLett.122.214502}
}

@article{Redor2021, title={{Experimental study of integrable turbulence in shallow water}}, volume={6}, ISSN={2469-990X}, url={https://link.aps.org/doi/10.1103/PhysRevFluids.6.124801}, DOI={10.1103/PhysRevFluids.6.124801}, number={12}, journal={Physical Review Fluids}, author={Redor, Ivan and Michallet, Hervé and Mordant, Nicolas and Barthélemy, Eric}, year={2021}, month=dec, pages={124801} }

@article{Randoux2016,
title = {{Nonlinear random optical waves: Integrable turbulence, rogue waves and intermittency}},
journal = {Physica D: Nonlinear Phenomena},
volume = {333},
pages = {323-335},
year = {2016},
note = {Dispersive Hydrodynamics},
issn = {0167-2789},
doi = {https://doi.org/10.1016/j.physd.2016.04.001},
url = {https://www.sciencedirect.com/science/article/pii/S0167278916301506},
author = {Stéphane Randoux and Pierre Walczak and Miguel Onorato and Pierre Suret}
}

@article{Agafontsev2015,
doi = {10.1088/0951-7715/28/8/2791},
url = {https://doi.org/10.1088/0951-7715/28/8/2791},
year = {2015},
month = {jul},
publisher = {IOP Publishing},
volume = {28},
number = {8},
pages = {2791},
author = {Agafontsev, D S and Zakharov, V E},
title = {{Integrable turbulence and formation of rogue waves}},
journal = {Nonlinearity}
}

@article{Agafontsev2016,
doi = {10.1088/0951-7715/29/11/3551},
url = {https://doi.org/10.1088/0951-7715/29/11/3551},
year = {2016},
month = {sep},
publisher = {IOP Publishing},
volume = {29},
number = {11},
pages = {3551},
author = {Agafontsev, D S and Zakharov, V E},
title = {{Integrable turbulence generated from modulational instability of cnoidal waves}},
journal = {Nonlinearity}
}

@article{Krasitskii1994, title={{On reduced equations in the Hamiltonian theory of weakly nonlinear surface waves}}, volume={272}, DOI={10.1017/S0022112094004350}, journal={Journal of Fluid Mechanics}, author={Krasitskii, Vladimir P.}, year={1994}, pages={1–20}}

@misc{Wu2025,
      title={{Validity condition of normal form transformation for the $\beta$-FPUT system}}, 
      author={Boyang Wu and Miguel Onorato and Zaher Hani and Yulin Pan},
      year={2025},
      eprint={2510.04831},
      archivePrefix={arXiv},
      primaryClass={math-ph},
      url={https://arxiv.org/abs/2510.04831}, 
}

@book{Nazarenko2011,
  title = {{Wave Turbulence}},
  ISBN = {9783642159428},
  ISSN = {1616-6361},
  url = {http://dx.doi.org/10.1007/978-3-642-15942-8},
  DOI = {10.1007/978-3-642-15942-8},
  journal = {Lecture Notes in Physics},
  publisher = {Springer Berlin Heidelberg},
  author = {Nazarenko,  Sergey},
  year = {2011}
}

@article{Zhang2022, title={{Numerical investigation of turbulence of surface gravity waves}}, volume={933}, DOI={10.1017/jfm.2021.1114}, journal={Journal of Fluid Mechanics}, author={Zhang, Zhou and Pan, Yulin}, year={2022}, pages={A58}}

@article{Dyachenko2004,
  title = {{Weak Turbulent Kolmogorov Spectrum for Surface Gravity Waves}},
  author = {Dyachenko, A. I. and Korotkevich, A. O. and Zakharov, V. E.},
  journal = {Phys. Rev. Lett.},
  volume = {92},
  issue = {13},
  pages = {134501},
  numpages = {4},
  year = {2004},
  month = {Apr},
  publisher = {American Physical Society},
  doi = {10.1103/PhysRevLett.92.134501},
  url = {https://link.aps.org/doi/10.1103/PhysRevLett.92.134501}
}

@article{Korotkevich2023,
  title = {{Inverse Cascade Spectrum of Gravity Waves in the Presence of a Condensate: A Direct Numerical Simulation}},
  author = {Korotkevich, Alexander O.},
  journal = {Phys. Rev. Lett.},
  volume = {130},
  issue = {26},
  pages = {264002},
  numpages = {6},
  year = {2023},
  month = {Jun},
  publisher = {American Physical Society},
  doi = {10.1103/PhysRevLett.130.264002},
  url = {https://link.aps.org/doi/10.1103/PhysRevLett.130.264002}
}

@article{Pelinovsky2006,
title = {{Numerical modeling of the KdV random wave field}},
journal = {European Journal of Mechanics - B/Fluids},
volume = {25},
number = {4},
pages = {425-434},
year = {2006},
issn = {0997-7546},
doi = {https://doi.org/10.1016/j.euromechflu.2005.11.001},
url = {https://www.sciencedirect.com/science/article/pii/S0997754605001007},
author = {Efim Pelinovsky and Anna {Sergeeva (Kokorina)}}
}

@article{Flamarion2024,
title = {{Nonlinear random wave fields within a Boussinesq system}},
journal = {Physics Letters A},
volume = {520},
pages = {129677},
year = {2024},
issn = {0375-9601},
doi = {https://doi.org/10.1016/j.physleta.2024.129677},
url = {https://www.sciencedirect.com/science/article/pii/S0375960124003712},
author = {Marcelo V. Flamarion and Efim Pelinovsky}
}

@article{Lvov2010, title={{Discrete and mesoscopic regimes of finite-size wave turbulence}}, volume={82}, rights={http://link.aps.org/licenses/aps-default-license}, ISSN={1539-3755, 1550-2376}, url={https://link.aps.org/doi/10.1103/PhysRevE.82.056322}, DOI={10.1103/PhysRevE.82.056322}, number={5}, journal={Physical Review E}, author={L’vov, V. S. and Nazarenko, S.}, year={2010}, month=nov, pages={056322}}

@article{Zhang2022b, title={{Forward and inverse cascades by exact resonances in surface gravity waves}}, volume={106}, ISSN={2470-0045, 2470-0053}, url={https://link.aps.org/doi/10.1103/PhysRevE.106.044213}, DOI={10.1103/PhysRevE.106.044213}, number={4}, journal={Physical Review E}, author={Zhang, Zhou and Pan, Yulin}, year={2022}, month=oct, pages={044213} }

@article{Hrabski2020, title={{Effect of discrete resonant manifold structure on discrete wave turbulence}}, volume={102}, ISSN={2470-0045, 2470-0053}, url={https://link.aps.org/doi/10.1103/PhysRevE.102.041101}, DOI={10.1103/PhysRevE.102.041101}, number={4}, journal={Physical Review E}, author={Hrabski, Alexander and Pan, Yulin}, year={2020},pages={041101} }

@article{Colleaux2025,
title = {A bound state attractor in optical turbulence},
journal = {Physica D: Nonlinear Phenomena},
volume = {477},
pages = {134687},
year = {2025},
issn = {0167-2789},
doi = {https://doi.org/10.1016/j.physd.2025.134687},
url = {https://www.sciencedirect.com/science/article/pii/S0167278925001630},
author = {Clément Colléaux and Jonathan Skipp and Sergey Nazarenko and Jason Laurie}
}

@article{Leduque2025,
  title = {{From deep to shallow water two-dimensional wave turbulence: Emergence of soliton gas}},
  author = {Leduque, Thibault and Kaczmarek, Maxime and Michallet, Herv\'e and Barth\'elemy, Eric and Mordant, Nicolas},
  journal = {Phys. Rev. Fluids},
  volume = {10},
  issue = {11},
  pages = {114801},
  numpages = {29},
  year = {2025},
  month = {Nov},
  publisher = {American Physical Society},
  doi = {10.1103/77r2-34v4},
  url = {https://link.aps.org/doi/10.1103/77r2-34v4}
}

@article{Chen2018, title={A Modified Newton Method for Nonlinear Eigenvalue Problems}, volume={8}, ISSN={2079-7362, 2079-7370}, url={https://global-sci.org/eajam/article/view/9365}, DOI={10.4208/eajam.100916.061117a}, number={1}, journal={East Asian Journal on Applied Mathematics}, author={Chen, Xiao-Ping and Dai, Hua}, year={2018}, pages={139–150} }

@article{Suret2024,
  title = {{Soliton gas: Theory, numerics, and experiments}},
  author = {Suret, Pierre and Randoux, Stephane and Gelash, Andrey and Agafontsev, Dmitry and Doyon, Benjamin and El, Gennady},
  journal = {Phys. Rev. E},
  volume = {109},
  issue = {6},
  pages = {061001},
  numpages = {35},
  year = {2024},
  month = {Jun},
  publisher = {American Physical Society},
  doi = {10.1103/PhysRevE.109.061001},
  url = {https://link.aps.org/doi/10.1103/PhysRevE.109.061001}
}

@article{Roberti2019,
  title = {{Early stage of integrable turbulence in the one-dimensional nonlinear Schr\"odinger equation: A semiclassical approach to statistics}},
  author = {Roberti, Giacomo and El, Gennady and Randoux, St\'ephane and Suret, Pierre},
  journal = {Phys. Rev. E},
  volume = {100},
  issue = {3},
  pages = {032212},
  numpages = {9},
  year = {2019},
  month = {Sep},
  publisher = {American Physical Society},
  doi = {10.1103/PhysRevE.100.032212},
  url = {https://link.aps.org/doi/10.1103/PhysRevE.100.032212}
}

@article{Bonnemain2022, title={{Generalized hydrodynamics of the KdV soliton gas}}, volume={55}, ISSN={1751-8113, 1751-8121}, url={https://iopscience.iop.org/article/10.1088/1751-8121/ac8253}, DOI={10.1088/1751-8121/ac8253}, abstractNote={Abstract
            We establish the explicit correspondence between the theory of soliton gases in classical integrable dispersive hydrodynamics, and generalized hydrodynamics (GHD), the hydrodynamic theory for many-body quantum and classical integrable systems. This is done by constructing the GHD description of the soliton gas for the Korteweg–de Vries equation. We further predict the exact form of the free energy density and flux, and of the static correlation matrices of conserved charges and currents, for the soliton gas. For this purpose, we identify the solitons’ statistics with that of classical particles, and confirm the resulting GHD static correlation matrices by numerical simulations of the soliton gas. Finally, we express conjectured dynamical correlation functions for the soliton gas by simply borrowing the GHD results. In principle, other conjectures are also immediately available, such as diffusion and large-deviation functions for fluctuations of soliton transport.}, number={37}, journal={Journal of Physics A: Mathematical and Theoretical}, author={Bonnemain, Thibault and Doyon, Benjamin and El, Gennady}, year={2022}, pages={374004} }

@article{El2021, title={{Soliton gas in integrable dispersive hydrodynamics}}, volume={2021}, ISSN={1742-5468}, url={https://iopscience.iop.org/article/10.1088/1742-5468/ac0f6d}, DOI={10.1088/1742-5468/ac0f6d}, abstractNote={Abstract
            We review the spectral theory of soliton gases in integrable dispersive hydrodynamic systems. We first present a phenomenological approach based on the consideration of phase shifts in pairwise soliton collisions and leading to the kinetic equation for a non-equilibrium soliton gas. Then, a more detailed theory is presented in which soliton gas dynamics are modelled by a thermodynamic type limit of modulated finite-gap spectral solutions of the Korteweg–de Vries and the focusing nonlinear Schrödinger (NLS) equations. For the focusing NLS equation the notions of soliton condensate and breather gas are introduced that are related to the phenomena of spontaneous modulational instability and the rogue wave formation. The integrability properties of the kinetic equation for soliton gas are discussed and some physically relevant solutions are presented and compared with direct numerical simulations of dispersive hydrodynamic systems.}, number={11}, journal={Journal of Statistical Mechanics: Theory and Experiment}, author={El, Gennady A}, year={2021}, month=nov, pages={114001} }

\end{document}